\documentclass{elsart}
\usepackage{natbib}

\usepackage{graphicx}
\usepackage{amssymb}

\newcommand{\bea}{\begin{eqnarray}}
\newcommand{\eea}{\end{eqnarray}}

\newcommand{\be}{\begin{equation}}
\newcommand{\ee}{\end{equation}}

\def\rhopack{\rho_{\textrm{\small pack}}}

\def\s{\textrm{s}}

\def\pN{\textrm{pN}}

\def\Nbp{N_\textrm{\small bp}}
\def\nm{\textrm{nm}}
\def\atm{\textrm{atm}}

\def\ang{\textrm{\AA}}
\def\phiphage{$\phi$29}
\def\lamphage{$\lambda$}
\def\mgion{$\textrm{Mg}^{2+}$}
\def\naion{$\textrm{Na}^{+}$}
\def\ds{d_\textrm{\small{s}}}
\def\Gbend{G_\textrm{\small{bend}}}
\def\Ghoop{G_\textrm{\small{hoop}}}
\def\Gent{G_\textrm{\small{ent}}}
\def\Gint{G_\textrm{\small{int}}}

\def\Gcap{G_\textrm{\small{cap}}}
\def\Gtot{G_\textrm{\small{tot}}}
\def\Rout{R_\textrm{\small{out}}}
\def\Ogenome{\Omega_\textrm{\small{genome}}}
\def\Ocapsid{\Omega_\textrm{\small{capsid}}}
\begin{document}
\begin{frontmatter}

\title{Forces During Bacteriophage DNA Packaging and Ejection}
\author[caltechE]{Prashant K. Purohit},
\author[caltechE]{Mandar M. Inamdar},
\author[caltechP]{Paul D. Grayson},
\author[caltechA]{Todd M. Squires},
\author[brandeis]{Jan\'{e} Kondev}, and
\author[caltechE]{Rob Phillips\corauthref{rob}}

\corauth[rob]{To whom correspondence should be addressed.}
\ead{phillips@aero.caltech.edu}
\thanks[caltechE]{Division of Engineering and Applied Science,
  California Institute of Technology}
\thanks[caltechP]{Department of Physics,
  California Institute of Technology}
\thanks[caltechA]{Department of Physics and Applied and Computational Mathematics,
  California Institute of Technology}
\thanks[brandeis]{Department of Physics, Brandeis University}

\begin{abstract}

The conjunction of insights from structural biology, solution
biochemistry, genetics and single molecule biophysics has provided a
renewed impetus for the construction of quantitative models of biological
processes. One area that has been a beneficiary of these experimental
techniques  is the study of viruses. In this
paper we describe how the insights obtained from such experiments can be
utilized to construct physical models of  processes in the viral life
cycle. We focus on dsDNA bacteriophages and show that the bending
elasticity of DNA and its electrostatics in solution can be combined to
determine the forces experienced during packaging and ejection of the viral genome.
Furthermore, we  quantitatively analyze the effect of
fluid viscosity and capsid expansion on the forces experienced during
packaging. Finally, we present a model for DNA ejection from
bacteriophages based on the hypothesis that the energy stored in the
tightly packed genome within the capsid leads to its forceful
ejection. The predictions
of our model can be tested through experiments \emph{in vitro} where  DNA ejection
is inhibited by the application of external osmotic pressure.

\end{abstract}

\end{frontmatter}


\nocite{alberts97}
\nocite{castelnovo03}
\nocite{gennes79}
\nocite{vries01}
\nocite{earnshaw80}
\nocite{evans01}
\nocite{kanamaru02}
\nocite{olson01}
\nocite{phillips01}
\nocite{riemer78}
\nocite{martin01}
\nocite{schnitzer00}
\nocite{simpson00}
\nocite{wikoff99}

\section{Introduction}

\begin{figure}
\centering
 \includegraphics[width = 10cm]{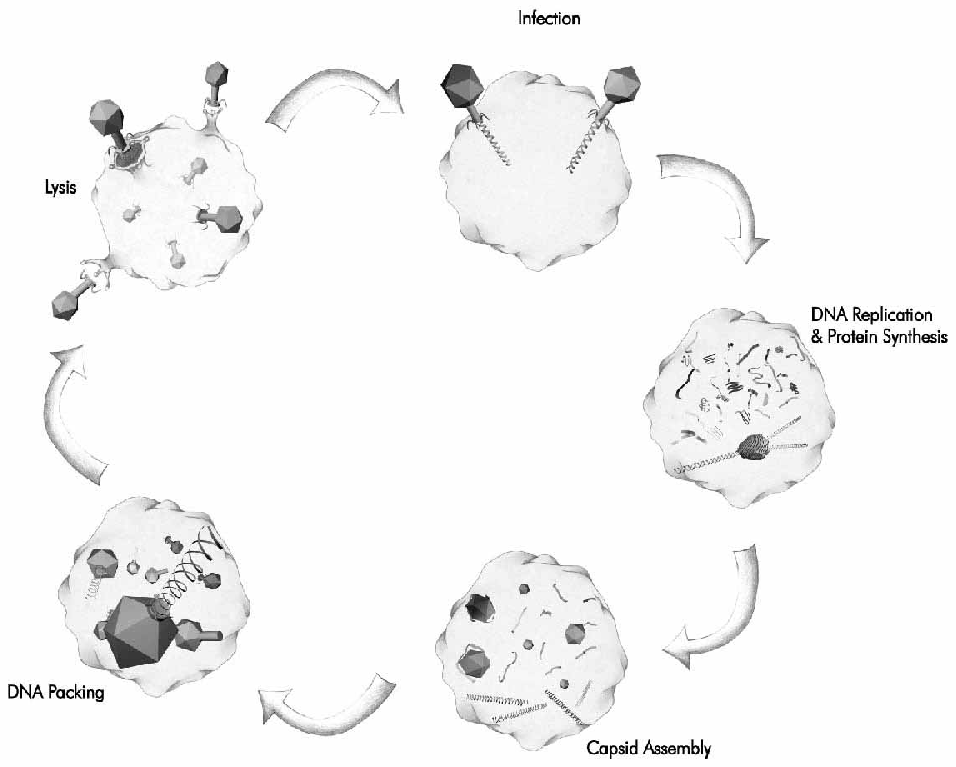}
 \caption{Life-cycle of a bacterial virus. The ejection of the genome
 into the host cell happens within a minute for phage like $\lambda$
 and  T4~\citep{novick88,letellier04}. The eclipse period (time
 between the viral adsorption and the first appearance of the progeny)
 is about 10-15 minutes~\citep{endy97}. The packaging of the genome
 into a single capsid takes about  5  minutes~\citep{smith01}. Lysis of the bacterial  cell is completed in less than an hour~\citep{endy97}.}
 \label{LifeCycle}
\end{figure}

Bacteriophage have served as a historical centerpiece in the
development of molecular biology.  For example, the classic
Hershey-Chase experiment~\citep{hershey52,echols01}
that establised nucleic acid to be 
the carrier of  the genetic blueprint was performed using bacteriophage T2. The biology
of bacteriophage $\lambda$ provided a fertile ground for the
development  of the understanding of gene
regulation~\citep{ptashne04}, while the study of virulent T phages (T1 to T7) paved the way for
many other advances such as the definition of a gene, the discovery of
mRNA, and
elucidation of the triplet code by genetic analysis~\citep{echols01}. One of the
central theses of the present paper is that phage are similarly poised
to serve as compelling model systems for quantitative analyses of
biological systems. Indeed, each of the stages of the viral life-cycle
(see Figure~\ref{LifeCycle}) can be subjected to physical analysis.

As shown in fig.~\ref{LifeCycle} a typical phage life-cycle consists of: adsorption, ejection,
genome replication, protein synthesis, self-assembly of capsid
proteins, genome packaging inside the capsid  and
lysis of the bacterial cell. A wealth of knowledge about various aspects
of these processes has been garnered over the last century with the
main focus being on
replication and protein-synthesis.
However, recent developments in the fields of x-ray crystallography, cryoelecton
microscopy, spectroscopy etc., have helped reveal the structural
properties of the viral components~\citep{baker99,leiman03} involved in ejection, assembly, and
packaging, while recent
experiments on {\phiphage}~\citep{smith01} and
\lamphage~\citep{evilevitch03} have helped quantify the forces
involved in the packaging and ejection processes, respectively. In this
paper we bring together these experimental insights
to formulate quantitative models for the packaging and ejection
processes.  The model is based on what we know about the structural
``parts list'' of a phage: the shape,
size, and strength of the capsid, the mechanical and electrostatic
properties of DNA, and high-resolution images that reveal the
structure of assembled phage particles.  Our goal is to create a
detailed picture of the forces implicated in DNA packaging and
ejection and, most importantly, make quantitative predictions that can be tested
experimentally.

The paper is organized as follows: in Section~\ref{lifecycle} we
examine dsDNA bacteriophage with the aim of assembling the relevant
insights needed to formulate quantitative models of packing and ejection. In Section~\ref{packaging} we develop the
model by examining the DNA packaging process in detail, and in
Section~\ref{ejection} we show that it can be used to explain the
DNA ejection process as well. The final section takes stock of a range
of quantitative predictions that can be made on the basis of the model
and suggests new experiments.

\section{Physical processes in the bacteriophage life cycle}\label{lifecycle}
Since the early measurements of virus sizes many experiments involving
viruses have been of a quantitative character. The emergence of
quantitive insights into viruses has come from many quarters including
electron microscopy, X-ray crystallography, single molecule biophysics
and
a large repertoire of classical methods in biochemistry and
genetics. For many viruses, these techniques have given us a detailed
picture of the structure and function of the entire viral parts list. Many viral genomes have been sequenced, and the
structure of many of the proteins coded for in the
genome \citep{baker99,Reddy01} have been solved. Though  it is
impossible to discuss all of these advances here, we
review some of the experimental insights that have informed our model building
efforts.

\subsection{Experimental Background}

DNA is highly compressed inside bacteriophage capsids, and the
resulting forces have important effects on the phage life cycle, as
revealed in several experiments.  One early experiment which shed light on the
possible role of the forces associated with packaged DNA is that of \citet{earnshaw77}, who characterized the tight packaging
of DNA in viral capsids by the distance $d_{s}$ between strands (around
2.8~nm in full capsids), and by \citet{feiss77}, who identified limits on the amount
of DNA that can be packaged into a {\lamphage} capsid.
\citeauthor{feiss77} found an upper limit barely above the wild-type
genome length, and suggested that adding more DNA makes the capsid
unstable.
\citet{rau84} made measurements on large volumes of non-viral DNA that
showed that $d_s$ values in the range of 2.5~nm to 3.0~nm correspond to a pressure
of several tens of atmospheres.

Since the experiments of \citeauthor{feiss77} and
\citeauthor{earnshaw77}, there have been a variety of experiments on
DNA packing, several of which stand out because of their relevance to
quantifying the internal
buildup of force during packaging.  \citet{shibata87} measured
the rate of packaging for phage T3, under various temperatures and
chemical conditions.  Through these experiments,
they determined that the packaging process was
reversible---one third of the genome was ejected back into solution
upon early interruption of packaging.
\citet{smith01} reinforced the
idea that a strong force builds up during packaging with real-time
single-phage packaging experiments.  They measured the rate of
DNA packaging while subjecting the DNA to various resisting forces,
quantifying the forces imposed by the packaging motor and the force resisting further
packaging as a result of the confined DNA.  Taken together,
these two experiments give us a picture of DNA packaging in which a
strong portal motor consumes ATP and in so doing pushes the DNA into
the capsid against an ever-increasing resistive force.


It is believed that the tightly wound DNA stores large energies
resulting in high forces, which in some cases could aid DNA ejection
into the host cell. Recently, an
experiment to
test this hypothesis was conducted by \citet{evilevitch03}, who
coerced {\lamphage} into ejecting its genome into a solution
containing polyethylene glycol (PEG) to create an external osmotic
pressure.  They found that various osmotic pressures of several tens
of atmospheres could halt the
ejection process resulting in fractional genome ejection. The
fractional ejection reflects a balance of forces between the inside and outside of
the capsid.  From this experiment it is concluded that forces are
still present during ejection at the same high levels as were
observed during packaging and in static capsids.

There are a variety of impressive {\it in vitro} experiments which
demonstrate pressure-driven ejection of the phage genome~\citep{novick88,bohm01}. 
For the {\it in vivo} case, ejection driven by internal force is only one of several
mechanisms that
have been hypothesized to participate in transferring
 the genome of bacteriophage into the host cell. Another mechanism is
suggested by \citet{molineux01} who, on the basis of a
wealth of experimental evidence, argues that the DNA of phage T7 is assisted
into the cell by DNA binding proteins. It is likely that different
bacteriophage use a combination of these two methods for
ejecting their DNA into the host cell.

Finally, recent cryo-electron microscopy (cryo-EM) and X-ray
crystallography studies of bacteriophage have revealed their detailed
internal structure and particularly, the ordered state of the packaged DNA.  \citet{cerritelli97} verified the tight
packing measured by \citeauthor{earnshaw77} and showed that the DNA is
apparently organized into circular rings within capsids. Other
structural experiments have revealed
the structure of components involved during ejection in T7
\citep{kanamaru02}, the assembly of bacteriophage
{\phiphage}\citep{tao98}, and structure of the packaging
motor in {\phiphage}\citep{simpson00}. This structural
information complements the single molecule measurements and will
guide us in the construction of a quantitative model of the
packaging and ejection processes.

\subsection{Orders of Magnitude in Bacteriophage Biophysics}
In the previous section we provided  background on some of the
experimental advances on bacteriophage which quantify the
packaging and ejection processes. It is the aim of the remainder of
this paper to
discuss the implications of these experiments in a more quantitative
way and to make predictions about the
phage life cycle that can be  tested experimentally.  Before describing
our models in precise terms, we first perform estimates of the orders
of magnitude of relevant physical properties involved in phage biophysics. Bacteriophage range in
size from a few tens of nanometers to several hundred
nanometers~\citep{baker99}. The capsids of most are
regular icosahedral structures a few tens of nanometers in size.
Table~\ref{tab:rufrho} gives an idea of the typical sizes of
bacteriophage, as well as some animal viruses for comparison. These
small containers house a genome which is several tens of microns
long, a feat that demands extremely efficient utilization of space.
In fact, a useful dimensionless quantity for characterizing the
packaging efficiency is
\begin{equation}
 \rhopack={\Ogenome \over \Ocapsid},
\end{equation}
where $\Ogenome$ is the volume of the genetic material and
$\Ocapsid$ is the volume of the capsid.
For double-stranded DNA bacteriophage, this result may be rewritten simply in
terms of the number of base pairs in
the phage DNA, $\Nbp$, using the approximation that DNA is a cylinder
of radius $1 \nm$ and length $0.34 \nm$ per base pair. Using these
approximations, $\rhopack$ can be rewritten as
\begin{equation}
 \rhopack = {0.34 \pi \Nbp \,  \over \Ocapsid} \,,
\end{equation}
making the calculation of $\rhopack$ straightforward. Note that
in this formula, $\Ocapsid$ is computed in units of $nm^3$.
\begin{table}
\centering
\begin{tabular}{|l|l|c|c|c|}     \hline
Virus type & Host type & genome length (kbp)
 & Diameter (\nm) & $\rhopack$ \\
\hline
Bacteriophage {T7} & Bacteria & 40 & 55 &  0.490 \\
Bacteriophage {\phiphage}\footnotemark[1] & Bacteria & 19.4 & 47 &  0.459 \\
Bacteriophage {T4} & Bacteria & 169 & 92 &  0.443 \\
Bacteriophage {\lamphage}\footnotemark[2] & Bacteria & 48.5 & 63 &
0.419 \\
Bacteriophage P22 & Bacteria & 41.7 & 63 & 0.319 \\
Herpes Simplex Virus 1 & Human & 152 & 125 & 0.159 \\
Human Adenovirus C  & Human & 36 & 80 & 0.143 \\
Smallpox Virus 1\footnotemark[3] & Human & 186 & 220 & 0.036 \\
Polyoma Virus SV40 & Human & $ 5.3 $ & $ \sim 50 $ & 0.083
\\
Mimivirus\footnotemark[4] & Amoeba & $\sim 800$ & $\sim 400$ & 0.026
\\
Papillomavirus BPV1  & Animal  & $7.9$ & $ 60 $ & 0.070
\\

\hline
\end{tabular}

\caption{Packaged volume fractions of some bacteriophage and
eukaryotic viruses. We have used the outer dimensions of the capsid
from \citet{baker99} in the calculation of $\rhopack$ since these
are more readily available than the dimensions of the empty space
inside the capsid.  The genome lengths are given, for most viruses, in
\citep{entrezgenomes}.  The DNA in bacteriophage is seen to be
significantly more tightly packed than the other viruses, revealing
the geometric origin of the large packing forces  associated with
bacteriophage.}\label{tab:rufrho}
\end{table}

\footnotetext[1]{\citep{tao98} Since the {\phiphage} capsids
  are aspherical, we use an average diameter that gives the correct
  volume.}
\footnotetext[2]{\citep{dokland93}}
\footnotetext[3]{\citep{whosmallpox} Since the smallpox
  particles are aspherical, we use an average diameter that gives the
  correct volume.}
\footnotetext[4]{\citep{lascola03}; this is the largest virus
  currently known.}

For the purposes of examining the significance of this parameter,
Table~\ref{tab:rufrho} shows $\rhopack$ for a number of different
viruses.  A trend that is evident in the table
is that viruses that infect bacteria are more tightly packed than the viruses
that infect eukaryotic cells.  A likely reason for this difference in
degree of genome compaction is the difference in infection strategies employed by the two
types of viruses.  While eukaryotic viruses are brought into a host
cell through processes in which both the genetic material and the
capsid are taken into the infected cell, bacteriophage typically attach to the
outside of the host and eject their DNA into the cytoplasm through a
small channel.  In order to transport their DNA quickly into the host,
which itself is pressurized at $\sim 3 \atm$~\citep{neidhardt96},
bacteriophage may power the ejection with a large internal pressure.
However, as  mentioned earlier, there is  experimental
evidence in the case of phage T7 that DNA binding proteins play a
role in DNA transport.  These experiments raise doubts about the
possibility of finding a single mechanism responsible for ejection
from all types of phage~\citep{molineux01}.

The bacteriophage life-cycle is a dynamic process and it is of
interest not only to consider the geometric parameters associated with
viral DNA, but to attend to the temporal scales that are involved as well.  The
first step in the cycle is the adsorption of the phage onto the host cell.
The frequency of this event depends on the abundance of available phage
particles and their hosts. Since about 50 to 300 new phage particles are
released by a single infection, the destruction of the host cell in a
culture proceeds
exponentially quickly once a cell is infected~\citep{young92}. After the phage has attached
itself to the host its genome is generally released on time scales
ranging from seconds to minutes~\citep{letellier04}. Probably concomitantly
the transcription and translocation machinery of the host cell is hijacked
and the production of phage proteins and factors required for replication
of its genome begins. The time between the adsorption of the phage and
the appearance of the first progeny capsids is usually on the order of
minutes~\citep{flint}.
This period is known as the `eclipse period' and it is the time required to
build up the concentration of the phage proteins to a level high enough to
initiate self-assembly of the capsid, tail and motor proteins that constitute
a mature phage particle. Self-assembly is a highly concentration dependent
process, but once it starts it proceeds rapidly to completion in a few seconds.

The second step, the packaging process, is completed in about 5--6
minutes \citep{smith01}.
The packaging rate is on the order of 100~bp/s in the initial stages but
it slows down as more of the genome is confined inside the capsid.
That is, the rate of packaging depends on the force opposing the
motor as it packages the genome. This internal force grows as the amount of
genome packaged increases, and the magnitude of the force depends on the solvent
conditions.  In fact, some biologically important multivalent ions, such as
spermidine, cause spontaneous DNA condensation resulting in much smaller
packaging forces~\citep{evilevitch04}. This will be demonstrated more clearly later in
this paper. After the progeny phage have been completely assembled,
enzymes lyse the host cell and a new generation of phage are released.
The process of adsorption to lysis is completed in less than an
hour~\citep{flint}. 

The objective of this section was to highlight some of the
opportunities for quantitative analysis provided by processes in the
viral life-cycle. Despite the wide range of interesting physical
process in the viral life-cycle, the remainder of this paper focusses
on the physical forces associated with packaged dsDNA and the
implications of these forces for the packing and ejection processes.


\section{The DNA Packaging Process}\label{packaging}

The model we invoke to examine the energetics of packaged DNA is
predicated upon two key physical effects: i) the elastic cost to bend
DNA so that it will fit in the viral capsid and ii) the interaction
energy which results from the proximity of nearby strands of DNA and
which derives from charges on both the DNA and in the surrounding
solution.

Our paper is very much inspired by previous theoretical work:
The work of~\citet*{riemer78} laid the foundation for subsequent
efforts on the energetics of packaged DNA by systematically examining
each of the possible contributions to the overall free energy
budget. \citet{kindt01} and \citet{tzlil03} estimated the forces in packaging of the phage \lamphage. \citet{arsuaga02}
computed conformations of DNA in phage~P4 using molecular mechanics
models, observing that the conformation of DNA within a capsid depends
on its volume, in agreement with our suggestion that the parameter
$\rhopack$ characterizes the extent of packing in phage.  \citet{odijk03} has discussed
several issues in bacteriophage packaging including the important problem of
deriving an expression for the electrostatic interactions between DNA strands
starting from the Poisson-Boltzmann equation with the added complication of
high density of packing and possibly non-hexagonal arrangement.
\citet{kindt01},\citet{purohit03} and \citet{tzlil03} do not derive these
expressions {\it a priori}; rather they use results obtained by
\citet{rau84}, \citet{parsegian86}, and \cite*{rau92} from osmotic
pressure experiments on DNA to characterize these interactions.  We have already
underlined the importance of electrostatic interactions in solutions as a
critical factor in the bacteriophage life cycle. More recently, \citet*{odijk03b}
analyze the possibility of a non-uniform density of DNA inside the capsid and
compute the size of regions within the capsid that are
void. \cite{marenduzzo03} study the effects of the finite thickness of
DNA on the forces experienced during packaging within a viral capsid. The aim
of this paper is to take models like those described above and to
systematically examine trends from one virus to another as well as for
mutants within a given phage type.

\subsection{Structural Models for the Packaged DNA}
In order to construct a quantitative model of DNA packaging and
ejection, it is necessary to specify the arrangement of the packaged DNA.
One of the earliest successful attempts at determining the structure of DNA packaged in
a phage capsid was an X-ray diffraction study by \citet{earnshaw77}
on P22 and some mutants of \lamphage. Their studies revealed both long-
range and short-range order of the encapsidated DNA. The evidence of long-range
order comes from the observation that the diffraction is modulated by a series
of ripples indicating that the phage head is uniformly filled. Short-range order
was indicated by a strong peak corresponding to a $25 \ang$ spacing. In an earlier study,
\citet{richards73} used electron microscopy to visualize gently disrupted
phage particles and found that the DNA close to the capsid boundary has a
circumferential orientation. These investigations together suggested a coaxial
spool like arrangement of the DNA inside the capsid. This model was further
corroborated by experiments of \citet{booy91} on HSV-1 and
\citet{cerritelli97} who used cryo-electron microscopy to obtain three-dimensional visualizations of the
packaged DNA in T7 phage. They consistently found a concentric ring-like geometry in both
wild-type and mutant versions of T7 with a spacing of around $25 \ang$ in
nearly-filled phage heads. However, none of these experiments shed much
light on the geometry of packaged DNA in the very early stages of packing.

A well ordered structure of the encapsidated DNA is also suggested by other
observations on DNA condensation in the presence of polyvalent cations
or other condensing agents such as methanol, ethanol or dilute solutions of
spermidine, PEG and other low molecular weight polymers
\citep[see][for an extensive overview]{gelbart00}. The strands in the
condensate are known to
have local hexagonal coordination, another piece of evidence in favor of the kind of short-range
order described above.  Simulations by \citet{kindt01} and
theoretical investigations by  \citet{odijk98} point in this direction as well.
Evidence in support of this hypothesis is also provided by the simulations of
\citet{arsuaga02} who show that the coaxial spool (or inverse spool)
is an energetically favorable configuration, particularly when the density
of packing is high as in bacteriophage. In light of these arguments we
{\it assume} that the DNA is arranged in columns of concentric hoops
starting from the inner wall of the capsid. Each of the hoops is surrounded
by six similar hoops (hexagonal arrangement) except those at the innermost
column and those touching the surface of the capsid. Though we have
assumed a limited class of geometries for the packaged DNA, namely
spool models, we note that such models are not necessarily the lowest
possible free energy structures~\citep{klug03}. Further, we remain
unclear as to the precise dynamic pathways that would orchestrate such
structures during packing and ejection, though simulations are
consistent with the dynamical development of such structures
\citep{arsuaga02, kindt01, lamarque04}


\subsection{Modeling the Free Energy of Packed DNA}
With a specific arrangement of DNA in hand, we can now compute the
free energy required to package that DNA.
Our efforts to write an expression for the free energy of the DNA packed inside
a phage capsid are inspired by the experimental insights
described above concerning the
configuration of the encapsidated DNA and its behavior in ionic environments.
We follow~\citep{riemer78},~\citep{odijk98} and~\citep{kindt01}
and break up the energetics of the encapsidated DNA into an elastic
term and a DNA-DNA interaction term:
\begin{equation}
 \Gtot(\ds,L) = \Gbend  + \Gint \,.
 \label{Gequation}
\end{equation}
We begin with the first term, by asserting that the bending energy in
an elastic fragment of length $L$ is given by
\begin{equation}
 G(L) = \frac{\xi_{p}k_{B}T}{2}\int_{0}^{L} \frac{dx}{R(s)^{2}} \,,
\end{equation}
where $\xi_{p}$ is the persistence length of DNA and $R(s)$ is the
radius of curvature
at the position with arc length parameter $s$. This simply reflects
the energetic cost $\Gbend = \xi_{p}k_{B}Tl/ 2R^{2}$ to bend an elastic rod of
length $l$ into a circular arc of radius of curvature $R$. 

The expression for the
bending energy is
considerably simplified by a structural insight provided by the
experiments described earlier of
\citet{earnshaw77}, \citet{booy91}, and \citet{cerritelli97} which showed
that the DNA is arranged in the capsid in a series of circular hoops starting
close to the surface of the capsid and winding inwards in concentric helices.
Therefore, we specialize this expression to a hoop of radius $R$
of length $l=2 \pi R$ and deduce that the bending energy in the hoop
is $\Ghoop = \pi \xi_{p}k_{B}T/R$. We ignore the pitch of the helix and think of
the encapsidated DNA as a series of hoops of different
radii~\citep{purohit03}. This leads to the following expression for
the total bending energy:
\begin{equation} \label{kutta}
 \Gbend(L) = \pi \xi_{p}k_{B}T \sum_{i} \frac{N(R_{i})}{R_{i}},
\end{equation}
where $N(R_{i})$ is the number of hoops of radius $R_{i}$ in the capsid.
Note that in neglecting the helical pitch we have also assumed that the
successive layers of hoops are parallel to each other. A more realistic
model would acknowledge that strands in successive layers form a criss-cross
pattern. We drop this complication in favor of the simple model used
above that
captures the essential physics.  The persistence length of DNA depends
on solvent conditions~\citep{smith92} and
also on the sequence of base-pairs~\citep{bednar95}. However, $\xi_{p} = 50$nm is an appropriate
number for the solvent conditions in the packaging experiment of
\citet{smith01} and the ejection experiment of \citet{evilevitch03}. We note here
that the size of many bacteriophage capsids is
on the order of a few tens of nanometers which is comparable to the
the persistence length of double stranded DNA itself. This hints that the bending
energy indeed constitutes a significant portion of the free energy of
the encapsidated DNA, though recent experiments~\citep{cloutier04}
reveal that bending DNA for small radius of curvature may not be as
costly as suggested by the linear elastic model used here.
Note that we have neglected twist and writhe as contributors to the free
energy very much in line with the work of
\citet{arsuaga02}, \citet{kindt01}, \citet{tzlil03}, and \citet{odijk03}. The twisting modulus of DNA is higher than its
bending modulus and the twist would be relaxed if the ends of the DNA are free
to rotate. However, there is no conclusive experimental data validating this
claim and more work is required on this point. There is a possibility that the DNA packaging motor is a rotational
device~\citep{simpson00} and investigations elaborating its mode of operation are expected
to shed more light on this problem.
Given this state of our knowledge we neglect
the twisting energy of the DNA and simplify the expression for the
bending energy by converting the discrete sum in eqn.(\ref{kutta}) to
an integral according to the prescription $\Sigma_{i}= {2\over
  \sqrt{3}d_{s}}\int dR$. In particular, we replace the sum of
eqn.(\ref{kutta}) with
\begin{equation} \label{eq:discrein}
 \Gbend(L) = \frac{2\pi \xi_{p} k_{B}T}{\sqrt{3}\ds}\int_{R}^{\Rout}
        \frac{N(R')}{R'}dR',
\end{equation}
where $\Rout$ is the radius of the inner surface of the capsid and the
length $L$ of packaged DNA is given by
\begin{equation} \label{eq:lenga}
 L = \frac{4 \pi}{\sqrt{3} \ds}\int_{R}^{\Rout} R'N(R') dR'.
\end{equation}
The factor of $\frac{\sqrt{3}}{2}\ds$ appears because it is observed
\citep{cerritelli97}  that
the DNA strands are arranged in a lattice with local hexagonal coordination
and spacing $\ds$. In other words, each strand has six nearest
neighbors except those near the surface of the capsid and the innermost
cylindrical space where there are three nearest neighbors on average. We
expect that the interaction of the DNA with the proteins of the capsid would
give rise to surface energy terms~\citep{tzlil03} in the expression for the total
free energy, but we neglect these terms in our analysis.\\

Though not written explicitly in
eqn.~\ref{Gequation}, entropy is another possible  contributor to the overall free energy budget of the
packaged DNA. We follow
Riemer and Bloomfield (1978) in showing that this is small in comparison to the other contributions to
the free energy arising from the elasticity of the DNA and its
interactions with itself. To see this we consider the extreme
case of free DNA that can explore a large set of
configurations in solution. This allows us to obtain an upper bound
for 
the entropic contribution to the free energy, since we remove the constraint imposed
by the tight packaging within the capsid. A fragment of DNA of
length \(L\) in a solvent can be modeled as a random walk on a three dimensional
lattice with lattice spacing \(2\xi_{p}\),where \(\xi_{p}\) is the persistence length~\cite{doibook}.
The total number
of links, \(N\), in the chain equals \(L/2\xi_{p}\) and the total number of
configurations of the chain on the lattice is \(z^{N}\), where \(z\)
is the co-ordination number of the chain ($6$ in this
case). Consequently, the
free-energy contribution due to the entropy is given by
\be
\Gent = {LkT\over2\xi_{p}}\ln 6 \approx {LkT\over \xi_{p}}.
\ee
A similar expression for the entropic contribution can also obtained
using Flory-Huggins type arguments following~\citet{riemer78}.
Since $\xi_{p} = 50 \rm nm$,
for bacteriophage $\phi29$ with $L = 6600 \rm nm$, $\Gent \approx
130k_{B}T$. This is two orders of magnitude smaller than the total energy of
$2\times 10^{4}k_{B}T$
obtained experimentally by~\citet{smith01} for
bacteriophage \(\phi29\). We therefore neglect the entropy of the DNA
in the rest of our calculations in this work. \\

\begin{figure}
\centering
 \includegraphics[width=10cm]{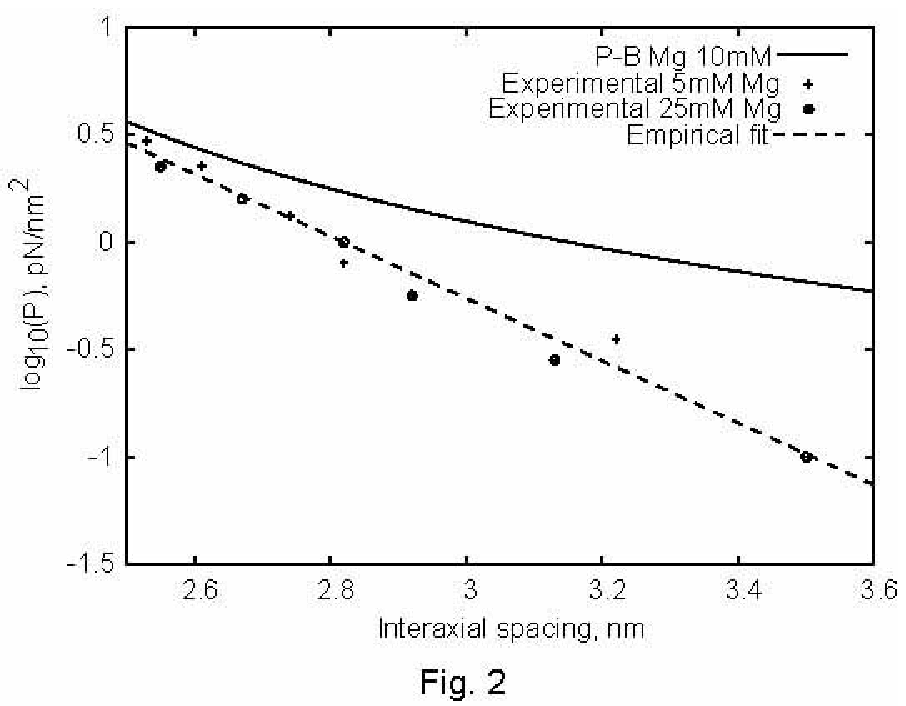}
 \caption{Pressure in a hexagonal lattice of DNA, according to
 experiment and Poisson-Boltzmann theory.  Experimental data points are
 from the data of \citet{rau84} for 25~mM and 5~mM {\mgion}
 concentrations. Our theoretical calculations follow from a discrete one-dimensional Poisson-Boltzmann solver, assuming
 cylindrical symmetry.  The free energy was calculated as a sum of the
 Shannon entropy of the ions and the electrostatic energy of ions and
 DNA, with the zero point for the potential set so that internal and
 external ionic concentrations were related by a Boltzmann factor.
 The theoretical predictions differ from the experimental points by a
 factor of ten, though the slopes are approximately correct and it is
 difficult to distinguish between the data for the two different
 concentrations. Also shown is
 a least-squares fit to the empirical datapoints, resulting in the
 parameters $c=0.30$~nm and $F_0= 1.2\times 10^{4}$~pN/nm$^2$.}
 \label{PBtest}
\end{figure}

We next turn to the DNA-DNA interaction energy $\Gint$.  DNA has
a backbone that is highly negatively charged. As a result there are large
energetic costs associated with bringing DNA fragments close together in
solution. These interactions can in turn be screened by counterions
which fill the space between DNA strands. From a theoretical perspective, it is
natural to treat these interactions by applying the
Poisson-Boltzmann approximation, in which charge is smoothly
distributed according to a Boltzmann distribution consistent with the
potential it produces~\citep{nelson03}. In this approximation, the
free energy is calculated from a sum of entropy (assuming the ions are locally an ideal
gas) and electrostatic energy.  Water is modeled as a continuous
dielectric.  In the limit of extremely close packing, ideal-gas
pressure dominates over the relatively constant electrostatic energy,
and the correct pressure is predicted. However, the experiments of
\citet{rau84} and \citet{parsegian86} show that this
approximation does not give the correct dependence of the pressure on
interstrand spacing as shown in Figure~\ref{PBtest}. The mismatch
between theory and experiment is likely due to
effects ignored by the Poisson-Boltzmann treatment, such as the discreteness of ions and of
water molecules.  However, detailed Monte-Carlo simulations in which
the ions are treated as discrete objects but water remains continuous
\citep{lyubartsev95} work remarkably well. Though our
treatment of the interaction energy, $\Gint$, will be based upon the
experimental data of \cite{rau84}, calculations like those of
\cite{lyubartsev95}
could be used to construct a ``first-principles''
model of the interactions, at least for monovalent and divalent
counterions. We rely on empirical data for this study but suggest the use of
simulated data if experimental data is not available.

The experiments of \citet{rau84} and
\citet*{rau92} provide an  empirical formula that relates osmotic
pressure $p$ to strand spacing $d_{s}$ in the range of two to four
nanometers as
\begin{equation}
 p(\ds) = F_{0}\exp(-\frac{\ds}{c}) \,.
\end{equation}
The value of the pressure is dependent on both the ion
concentration and its charge. When
working with this formula, it is important to remember that a large
change in $F_{0}$ can be mostly compensated by a small change in
$c$, since most data points are in a small region far from the
$p$-axis.  Hence, even though $c$ is relatively constant, we
state the value used for both $c$ and $F_0$ together to avoid
confusion.
Figure~\ref{PBtest} shows the experimental data for a solution
containing 5 and 25~mM MgCl$_2$ at 298~K, in which measurements reveal
$c=0.30$~nm and $F_0= 1.2 \times 10^{4}$~pN/nm$^2$.   These values should be appropriate
for use whenever {\mgion} ions are the dominant species and have a
concentration of 5--25~mM, conditions satisfied in solutions commonly
used in phage experiments, including, for example, SM~buffer and the
TM~buffer used in \cite{evilevitch03}.  Unfortunately, the
buffer used in \citet{smith01}, includes enough sodium to
suggest that the effect of {\naion} must be considered in addition to
{\mgion}.
The solutions {\it in vivo} are far more
complex and $F_{0}$ is difficult to determine. Even solutions {\it in vitro}
contain different concentrations of ions and determining $F_{0}$ for each
one of them through experiments would be impractical.

The data on the measured pressure in terms of $\ds$ can be
used to deduce the
functional form of energy stored in the electrostatic interactions. We do
not go through the details here but refer the reader to \citet{purohit03}
for the full calculation. The calculation rests on the assumption of
a pair potential interaction among $N$ parallel strands of length $l$ each
packed in a hexagonal array with a spacing $\ds$. The total interaction
energy of this arrangement is
\begin{equation}
 \Gint(L,\ds) = \sqrt{3}F_{0}(c^{2}+c\ds)L \exp(-\frac{\ds}{c}),
\end{equation}
where $L=Nl$ is the total length of the strands. This is the expression
for the interaction energy when the ions in the ambient solution are monovalent
or divalent cations. In this regime the interaction between the
strands is entirely repulsive and these interactions decay  as the strands are moved farther apart.
For trivalent and tetravalent ions the physics is quite different. DNA in such
a solution can condense into hexagonally packed tori
\citep{raspaud98}. In this regime there is a preferred spacing
$d_{0}$ between the strands, and the measurements are well fit by,
\begin{equation}
  \Gint(L,\ds) = \sqrt{3}F_{0}L\Big[(c^{2}+c\ds)\exp(\frac{d_{0}-\ds}{c})
                     - (c^{2}+cd_{0}) - \frac{1}{2}(d_{0}^{2}-\ds^{2})\Big],
\end{equation}
where $c=0.14 \nm$ and $d_{0} = 2.8 \nm$. For $\ds < d_{0}$ the interaction
is strongly repulsive and for $\ds > d_{0}$ it is attractive. This
expression is a good representation of the free energy of interactions
between the DNA molecules for spacings $d_{s}$ less than or equal to
the preferred value $d_{0}$~\citep{rau92}. In viral packaging we
encounter interaxial spacings in exactly this range and hence we will
use this free energy to  study the effects of repulsive-attractive
interactions on encapsidated DNA.

Given that we have now examined the separate contributions arising
from DNA bending, entropy and interaction terms, we now write the free
energy of the encapsidated DNA
in the repulsive regime,
\begin{equation}\label{total_energy}
 \Gtot(L,\ds) =  \underbrace{\frac{2\pi\xi_{p}k_{B}T}{\sqrt{3}\ds}\int_{R}^{\Rout}
               \frac{N(R')}{R'}dR'}_{\mbox{Bending}}
               + \underbrace{\sqrt{3}F_{0}(c^{2}+c\ds)L\exp(-\frac{\ds}{c}) \,.}_{\mbox{Interaction}}
\end{equation}
An analogous formula holds for the repulsive-attractive regime.
Note that this expression reports the free energy of the inverse spool
configurations when a length $L$  has been
packaged by relating, $R$ to $L$ via~(\ref{eq:lenga}).
We will now show that the spacing between the strands
varies in a systematic way during the packing process reflecting the
competition between bending and interaction terms.

\subsection{DNA Spacing in Packed Capsids}

\begin{figure}
\centering
 \includegraphics[]{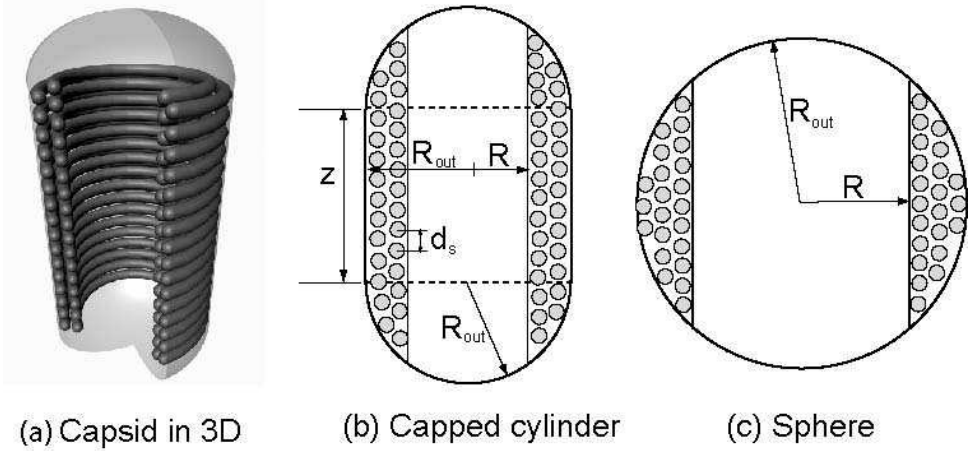}
 \caption{Idealized geometries of viral capsids.}
 \label{fig:pilla}
\end{figure}
Our model makes a concrete prediction for the free energy $\Gtot$ of packaged
DNA in any phage and for a wide range of solution conditions.  In order to find $\Gtot$ for a particular phage, we
minimize eqn.(\ref{total_energy}) by varying $\ds$, under the
constraint that $L$ given by (\ref{eq:lenga}) is equal to the
length of DNA already packaged. The expressions
for $N(R')$ will differ for different capsid geometries.
Most capsids are icosahedral and we idealize them as spheres.
Some capsids, e.g., {\phiphage}, have a ``waist''. We idealize them as cylinders
with hemispherical caps. Eventually, we will find that the geometry
does not affect the overall free energy of packing
as long as the internal volume of the idealization is the same for each
geometry, once again reflecting the importance of the parameter $\rho_{\rm pack}$, introduced
earlier.
Before we specialize to particular geometries (see fig.~\ref{fig:pilla})we observe that
$N(R') = \frac{z(R')}{\ds}$ where $z(R')$ is the height of a column
of hoops of DNA situated at radius $R'$. Using this fact and differentiating
(\ref{eq:lenga}) with respect to $\ds$, while holding $L$ 
constant, gives us $\frac{dR}{d d_{s}} = -\frac{\sqrt{3}L d_{s}}{2\pi
  Rz(R)}$. Minimizing $\Gtot$ with respect to the interstrand spacing $d_{s}$ gives,
\begin{equation} \label{eq:tosolve}
 \sqrt{3} F_{0}\exp(-\ds/c) =\frac{\xi_{p} k_{B}T}{R^{2}\ds^{2}}
 -\frac{\xi_{p} k_B T}{\ds^{2}} \frac{\int_R^{\Rout}
  \frac{z(R')}{R'} dR'}{\int_{R}^{\Rout} R' z(R') dR'}.
\end{equation}
This equation represents a balance between the bending energy terms and
the interaction terms. Note that if the size of the capsid is fixed then
longer lengths of packed DNA imply smaller radii of curvature for the hoops
since the strands want to be as far away as possible from each other
for the case in which the interaction between the adjacent strands is
repulsive. On the other hand, smaller radii of curvature eventually lead to
large bending energy costs resulting in a trade-off that is captured
mathematically in eqn. (\ref{eq:tosolve}).

We now specialize this result to particular geometries
(i.e. particular choices of $z(R')$). For a sphere, $z(R') =
2\sqrt{\Rout^{2}-R'^{2}}$, and the equation which determines the
optimal  $\ds$ reads
\bea 
 \sqrt{3} F_{0}\exp(-\ds/c)  &=& \frac{\xi_{p} k_{B}T}{R^{2}\ds^{2}}
                      + \frac{3\xi_p k_B T}{\ds^{2}}
                        \Big(\frac{1}{\Rout^{2} - R^{2}}\nonumber \\
                      &+& \frac{\Rout}{(\Rout^{2} - R^{2})^{3/2}}
                        \log\big(\frac{\Rout - \sqrt{\Rout^{2}
                        - R^{2}}}{R}\big)\Big).
\eea
For a cylinder with hemispherical caps, $z(R') = h + 2\sqrt{\Rout^{2}-R'^{2}}$,
where $h$ is the height of the waist portion, and  $\ds$ is
the solution of the following equation:
\bea
  \sqrt{3} F_{0}\exp(-\ds/c)  &=& \frac{\xi_{p}
                      k_{B}T}{R^{2}\ds^{2}}\nonumber \\
                      &+& \frac{\xi_p k_B T}{\ds^{2}}
                        \frac{\Big(h\log(\frac{\Rout}{R})
                              - \sqrt{\Rout^{2} - R^{2}}
                              - \log\big(\frac{\Rout - \sqrt{\Rout^{2}
                              - R^{2}}}{R}\big)\Big)}
                             {\frac{h}{2}\sqrt{\Rout^{2}-R^{2}}
                             + \frac{1}{3}(\Rout^{2}-R^{2})^{3/2}}\label{eq:capcyl}.
\eea
\begin{figure}
\centering
 \includegraphics{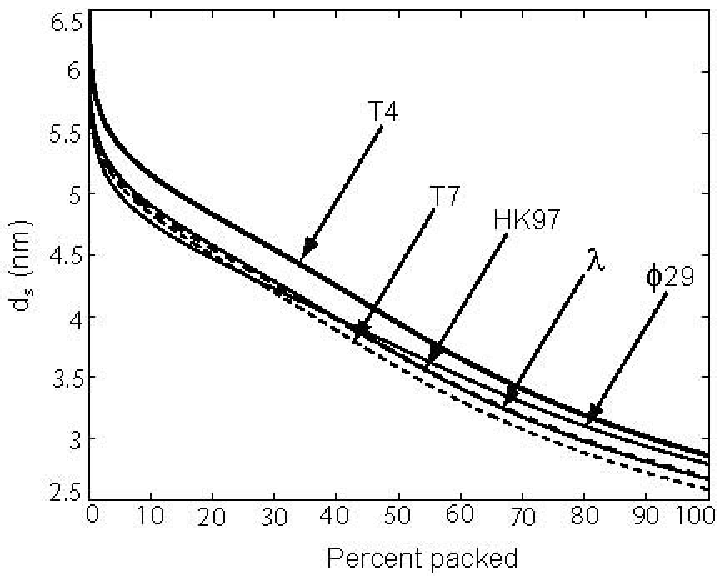}
 \caption{The spacing $\ds$ in five different phage under fully repulsive
 conditions with $F_{0}=2.3 \times 10^{5}$pN/nm$^{2}$ and $c=0.27$nm. These
 values of $F_{0}$ and $c$ result in
 the best visual fit to the data in the experiment on {\phiphage} by
 \citet{smith01}($5$mM $\rm MgCl_{2}$ and $50$mM NaCl). The graphs show $\ds$ monotonically
 decreasing as more of the genome is packaged. T7 is the most tightly packed while
 T4 is most loosely packed. $\lambda$ and HK97 show an almost identical history
 of $\ds$ vs. fraction of genome packed, since the two
 are closely related structurally.}
 \label{fig:dspac}
\end{figure}
Figure~\ref{fig:dspac} shows $\ds$ as a function of the
fraction
of the genome packed for five different phage. We assume that all of them
are packaged under the same (repulsive) conditions, with $F_{0} =
2.3 \times 10^{5}$pN/nm$^{2}$
and $c=0.27$nm as in~\citet{purohit03} corresponding to the best
visual fit to the data in the experiments of~\citet{smith01} which
was performed in a solution containing $5$mM $\rm MgCl_{2}$ and $50$mM
NaCl. Table~\ref{tab:dime} gives the details of the geometry of
the phage used in the calculation.
\begin{table}
\centering
\begin{tabular}{|l|l|c|c|c|c|} \hline
Phage type & Model & $\Rout$ (nm) & $h$ (nm) & Genome
& $\rhopack$ \\
 & Geometry& & & length (nm)& \\\hline \hline
T4~\citep{iwasaki00}         & Capped cylinder & 39.8  & 29.0 & 57424  & 0.442
\\ \hline
T7$\dagger$~\citep{cerritelli97}  & Sphere       & 26.6  & 0    & 13579  & 0.541
\\ \hline
{\phiphage}~\citep{tao98} & Capped cylinder & 19.4  & 12.0 & 6584   & 0.461
\\ \hline
HK97~\citep{lata00}      & Sphere          & 27.2  & 0    & 13509  & 0.503
\\ \hline
$\lambda$~\citep{baker99} & Sphere          & 29.0  & 0    & 16491  & 0.507
\\ \hline
\end{tabular}
\caption{Idealized geometries of bacteriophage. The radius and height
are determined by using the volume available to the DNA. They have been calculated
from experimental data about the geometry of capsids from several sources
(in parentheses above). $\rhopack$ for some phage
in this table is higher than
corresponding values in table~\ref{tab:rufrho} since this table uses internal
volumes whereas table~\ref{tab:rufrho} uses the outer dimensions that are more
readily available. $\dagger$ The T7 phage is unusual in
that a part of its cylindrical tail (radius $9.5$nm, height $28.5$nm)
protrudes into the empty space within the spherical capsid. The space occupied
by this tail is not available to the DNA and we exclude it to get an effective
radius.} \label{tab:dime}
\end{table}

\begin{figure}
\centering
 \includegraphics{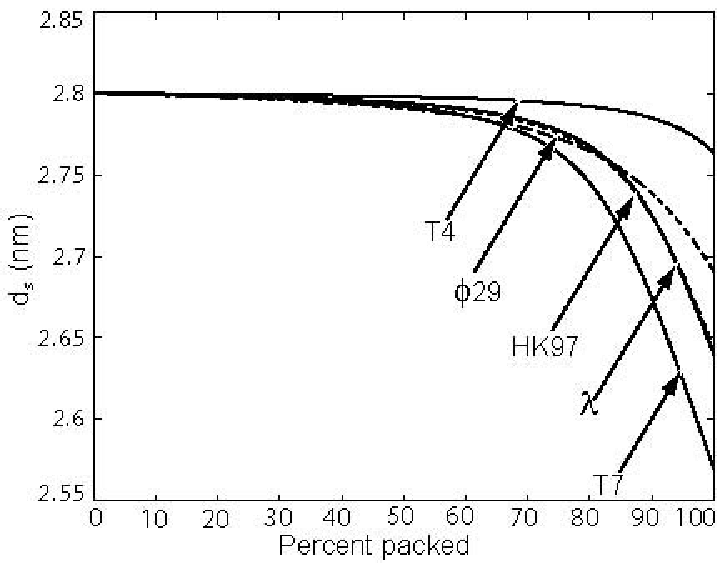}
 \caption{The spacing $\ds$ under repulsive-attractive conditions. We
  use $F_{0} = 0.5$ pN/nm$^{2}$, $d_{0} = 2.8$nm and $c =
  0.14$nm. This corresponds to a solution containing $5$mM $\rm
  Co(NH_{3})_{6}Cl_{3}$, $0.1 \rm M$ NaCl, $10 \rm mM$ TrisCl~\citep{kindt01,rau92}. The
  spacing
  remains at the preferred value of $2.8$nm for most of the packaging process,
  except in the end when volumetric constraints lead to smaller spacings as a
  consequence of   high energy costs for maintaining this $\ds$.}
 \label{fig:attspac}
\end{figure}

As noted earlier, a sufficient concentration of polyvalent cations
suffices to induce effective attraction between adjacent DNA strands.
Under such repulsive-attractive conditions the left hand side of
eqn.(\ref{eq:tosolve}) changes and we get
\begin{equation} \label{eq:repatr}
 \sqrt{3} F_{0}(\exp(\frac{d_{0}-\ds}{c}) - 1)
 =\frac{\xi_{p} k_{B}T}{R^{2}\ds^{2}}
 -\frac{\xi_{p} k_B T}{\ds^{2}} \frac{\int_R^{\Rout}
  \frac{z(R')}{R'} dR'}{\int_{R}^{\Rout} R' z(R') dR'}.
\end{equation}
Here too, the idea is to solve for the optimal $\ds$ at each packaged
length $L$. The results with the repulsive-attractive potential for different phage are
shown in Figure~\ref{fig:attspac}. We note the similarity in the observed
trends with the results of \citet{kindt01} who also
study the packaging process with a repulsive-attractive potential. In the presence of surface energy
the encapsidated DNA assumes a toroidal configuration in the early stages
of packing \citep{kindt01} much as it would have in solution in
the absence of a capsid \citep{raspaud98}. At the later stages
of packing the inverse spool is the more energetically favorable geometry and
the calculations of \citeauthor{kindt01} show a smooth transition between these
configurations.  However, we assume that the inverse spool is the optimal geometry
throughout the packaging process since we do not have surface energy
terms. Note also that the trends in figure~\ref{fig:attspac} are quite different
from those with
the fully-repulsive conditions, figure~\ref{fig:dspac}. Most importantly, in the early stages of
packing under repulsive-attractive conditions the DNA strands tend to be at
the preferred spacing of $2.8$nm~\citep{kindt01, tzlil03}. Volumetric constraints at the later stages
of packing result in tighter packing and $\ds$ decreases
to values lower than the preferred $2.8$nm. Under fully-repulsive
interactions the interstrand spacing decreases monotonically as more DNA
is packaged.

\begin{figure}
\includegraphics{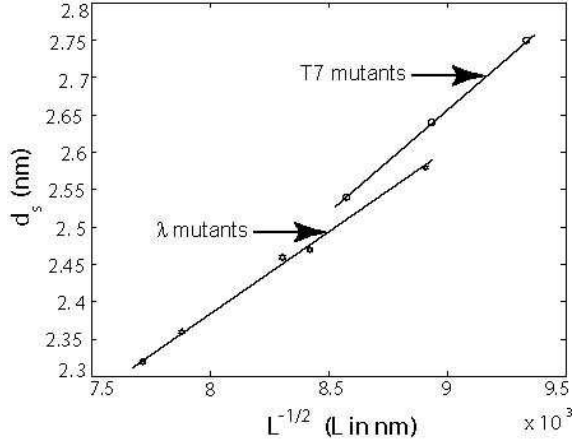}
\centering
\caption{Comparison of measured spacings $\ds$ with $\frac{1}{\sqrt{L}}$ scaling law.
The circles correspond to interstrand spacing in T7 obtained by
\citet{cerritelli97}. The stars represent data from \citet{earnshaw77}
for mutants of $\lambda$-phage. We fit straight lines to both
these data sets to show that the spacing scales with the inverse square root
of the packaged length in capsids that are nearly full.}
\label{fig:rootL}
\end{figure}

For many bacteriophage, even were they to be packaged under conditions
in which there is an effective attraction between adjacent DNA
segments the genome would not be fit into the capsid if the interaxial
spacing is at the preferred value of $2.8$nm. This would result in 
repulsive interactions between adjacent DNA segments in the terminal part of the packaging process. 
Interestingly, in these circumstances, $\ds$
can be estimated using strictly geometric
arguments. In particular, we equate the total volume available in the capsid
with the volume of the packaged DNA. In particular, this implies
\begin{equation}
 V_{cap} = \frac{\sqrt{3}}{2}\ds^{2} L,
\end{equation}
where $L$ is the length of the packaged DNA. This in turn implies
\begin{equation}
 \ds = \sqrt{\frac{2V_{cap}}{\sqrt{3}}} \frac{1}{\sqrt{L}},
\end{equation}
or more generally $\ds \propto 1/\sqrt{L}$. In deriving this expression
we have assumed that the volume of the cylindrical void (unoccupied by the
DNA)~\citep{arsuaga02, kindt01, odijk98} in the middle of the capsid is negligible in comparison to the volume
of the capsid.
The predicted scaling of $\ds$ with DNA length $L$ above can be compared to
several different experiments as shown in Figure~\ref{fig:rootL}. In
particular, the spacing has been measured both in $\lambda$ mutants
\citep{earnshaw77} and in the T7 bacteriophage
\citep{cerritelli97}. As seen in the figure,
the scaling suggested by the model appears to provide a satisfactory
description of the measured trends. More generally, note that the
spacing of the packaged DNA is one of the key points of contact
between our theory and experiment. In particular, both
fig.~\ref{fig:dspac} and fig.~\ref{fig:attspac} are \emph{predictive} in that
they suggest how the spacing $\ds$ varies from one phage to another
for different solution conditions.

\subsection{Forces During DNA Packing}

The phage genome is subject to various resistive forces during the
packaging process. As discussed earlier, these forces depend on the
size and geometry of the capsid, solvent conditions and the size of
the genome. In this section we calculate
those forces using the expressions for the elastic and the interaction
energy derived in the previous section.

\subsubsection{Forces from DNA confinement}
\begin{figure}
\centering
\includegraphics{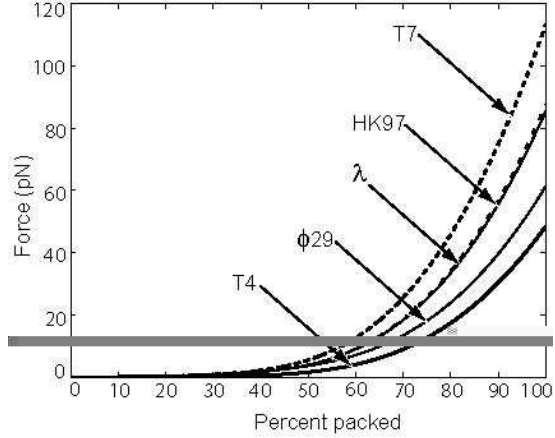}
\caption{Comparison of forces during DNA packing process for different phage
 under fully repulsive conditions. T7 requires the largest force to package
 since it is most densely packed. T4 is at the other end of the spectrum
 requiring the smallest force. The data above corresponds to $F_{0} =
 2.3\times 10^{5}$
 pN/nm$^{2}$ and $c=0.27 \rm nm$ obtained by a visual fit to the
 data of~\cite{smith01}, who conducted the packaging experiment for
 phage {\phiphage} in a solution containing $5\rm{mM}$ $\rm MgCl_{2}$ and $50 \rm mM$ NaCl.}
\label{fig:Trends}
\end{figure}

\begin{figure}
\centering
\includegraphics{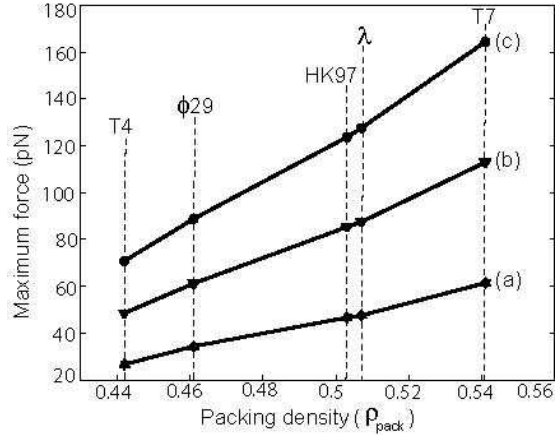}
\caption{Maximum resistive force in different phage under three
 different repulsive conditions. (a) corresponds to $F_{0} = 1.1\times
 10^{5}$ pN/nm$^{2}$ and $c = 0.27$nm,
 (b) corresponds to $F_{0} = 2.3\times 10^{5}$ pN/nm$^{2}$ and $c=0.27$nm and
 (c) corresponds to $F_{0} = 3.3\times 10^{5}$ pN/nm$^{2}$ and $c=0.27$nm. The forces increase with the
 packing density $\rhopack$ and also with increasing $F_{0}$.}
\label{fig:cooldude}
\end{figure}

\begin{figure}
\centering
\includegraphics{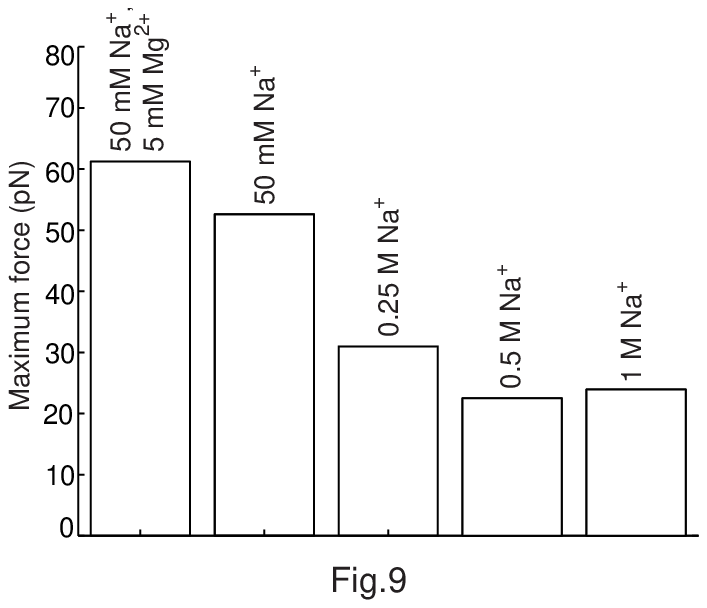}
\caption{Maximum resistive force in {\phiphage}  for different salt
 concentrations. The maximum force increases as the salt concentration
 decreases since the DNA interstrand repulsion becomes larger as the
 solution becomes more dilute. The
 values of $F_{0}$ and $c$ for the salt solutions used in this figure
 were obtained from fits to the data of~\cite{rau84}. }
\label{fig:bachha}
\end{figure}

Using the model described above, it is possible to evaluate the forces resisting
packaging for different phage. To obtain the force we differentiate the total free energy, $\Gtot$,
with respect to the packaged length $L$. Since the dependence of the spacing
$\ds$ on $L$ is already known we substitute it into the expression for $\Gtot$
rendering it  a function of $L$ alone. We then carry
out the differentiation and obtain the following expression for the force:
\begin{equation}
 F(\ds(L),L) = -\frac{d\Gtot}{dL} = \sqrt{3}F_{0}\exp(c^{2}+c\ds)
                                    + \frac{\xi_{p}k_{B}T}{2R^{2}},
\end{equation}
where we assume that the radius $R$ and the spacing $\ds$ are known
functions of $L$.
In Figure~\ref{fig:Trends}, we show the result of a series of such calculations
for different phage, all done assuming ionic conditions such as
those used in the experiments of \citet{smith01}. The value of $F_{0}
\mbox{ and } c$ corresponding to these conditions were determined by
a fit to the data of~\cite{smith01} for bacteriophage
$\Phi29$. In principle more accurate values of these parameters can be
obtained by a least-squares fit. The key point of
Figure~\ref{fig:Trends} is to illustrate the trends across different phage with
particular reference to the way in which the maximum packaging force scales
with $\rhopack$. This is shown more clearly in Figure~\ref{fig:cooldude}. We
observe that the
maximum resistive force scales roughly linearly with the packing density
$\rhopack$ across different phage of varied shapes. We have also
plotted the maximum resistive force for {\phiphage} in
Figure~\ref{fig:bachha} for different salt concentrations of the
ambient solution. The repulsive interactions between the DNA
strands grow progressively larger as the concentration of the $\rm
Na^{+}$ ions in solution decreases. This results in larger packaging
forces as can be seen in Figure~\ref{fig:bachha}.

\begin{figure}
\centering
\includegraphics{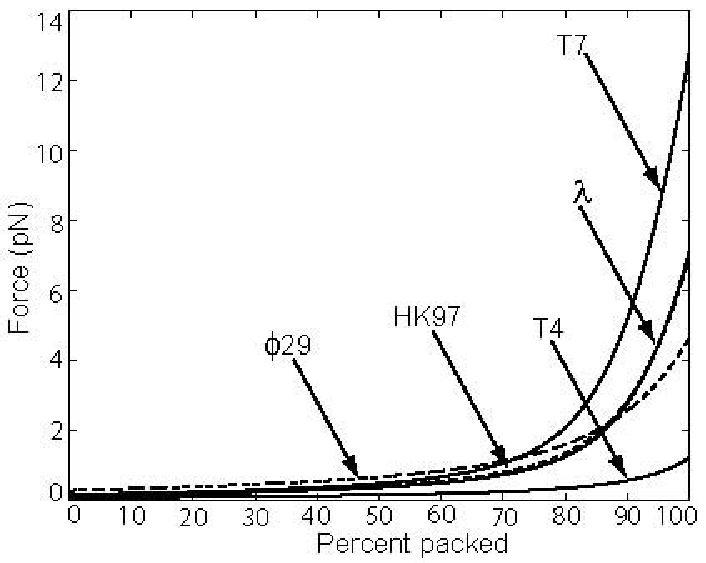}
\caption{Comparison of forces during DNA packing process for different phage
under repulsive-attractive conditions with $F_{0} = 0.5$ pN/nm$^{2}$, $d_{0} = 2.8$nm and $c =
  0.14$nm. This corresponds to a solution containing $5$mM $\rm
  Co(NH_{3})_{6}Cl_{3}$, $0.1 \rm M$ NaCl, $10 \rm mM$ TrisCl~\citep{kindt01,rau92}. The trends seen here are no different
from those in the fully repulsive conditions - T7 requires large forces and
T4 requires small forces for packaging. The maximum force, however, is
significantly smaller than that seen for fully-repulsive conditions.}
\label{fig:attfor}
\end{figure}

We have also obtained the force required for packaging under repulsive-attractive
interactions. The results can be seen in Figure~\ref{fig:attfor}. The forces
are considerably smaller when the strands are at the preferred spacing
$d_{0} = 2.8$nm. As with the DNA spacing discussed earlier the results
in Figures~\ref{fig:Trends} and \ref{fig:cooldude} are
\emph{predictive} and suggest a wide range of new experiments using
both different solution conditions and different phage.

\begin{figure}
 \centering
 \includegraphics{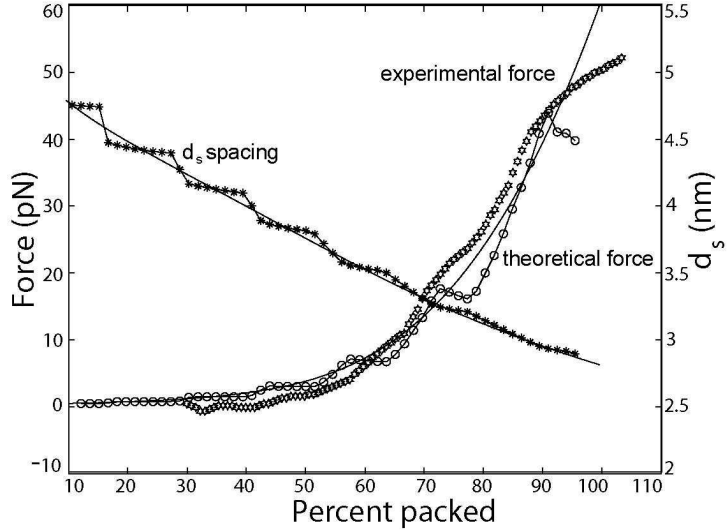}
 \caption{Force and interaxial spacing as functions of the amount of DNA
 packed in bacteriophage {\phiphage}. The
 hexagons correspond to the experimental data of \citet{smith01}.
 The thick lines are results of the continuum model and the circles connected
 with the thin line are obtained from the discrete model.}
 \label{fig:discphi}
\end{figure}
The history of force as a function of packaged length obtained above
rested on the assumption that the integral approximation adopted in
eqn.(\ref{eq:discrein}) is an accurate representation of the bending
free energy. In principle the optimal spacing $\ds(L)$ and the
resistive force $F(L)$ can be obtained by minimizing the free energy
without resorting to the integral approximation. This has been carried
out in~\cite{purohit03b} and the results are shown in
Figure~\ref{fig:discphi}. There are several distinctive
features to be noted in this figure. Foremost among them is that the
curves for the history of resistive force and interaxial spacing as
a function of the length of DNA packed are not monotonic, unlike the curves obtained
from the integral (continuum) models. The discrete steps 
represent the addition of a new stack of hoops, during the packaging
reaction. More importantly, discreteness of packing implies that at specific
lengths packaged there will be sharp changes of the spacing~$\ds$ due to
an increase in the number of hoops per layer. These events might be a
mechanical signal of the packaging configuration. In particular, the
observation of such steps in an experiment would be evidence in
favor of our quasi-static picture of the dynamics of packaging.



\subsubsection{Forces due to viscous dissipation}

Thus far we have calculated the free energy stored in the compressed
DNA within a bacteriophage capsid, which represents the total amount
of work that the motor must do to package the entire genome.
We have not yet considered irreversible work that may result
from the high rate of packaging.  An obvious source of irreversible
work is viscous dissipation in the fluid.  Here we demonstrate that, in fact, 
such forces are negligible, as speculated by~\citet{smith01}. 

We identify four sources of viscous dissipation during
packaging: drag on (i) the capsid and 
(ii) the unpackaged genome as each is pulled through the fluid, (iii)
viscous dissipation
within the sheath (as might be important during ejection), and (iv) dissipation as fluid is extruded through
capsid pores.
We calculate an upper bound on each to provide
an upper bound on fluid dissipation in general, and conclude that such
forces
are negligible.  Note, however, that we assume water to behave as a
continuum
material throughout this analysis.  Non-continuum effects, which could
affect
our conclusions, are not treated here.  In what follows, we use values
from Table
\ref{tab:parameters}.

\begin{table}
\begin{center}
\begin{tabular}{|c|l|c|}
\hline
$R$     & Capsid radius & 30 nm\\
$t_c$ & Capsid thickness & 2 nm \\
$R_p$ & Pore radius & 2 nm \\
$N$ & Number of pores & 10\\
$X$     & Sheath length         & 50 nm         \\
$D$     & Inner  sheath radius  & 1.3 nm        \\
$d$     & DNA radius            & 1.0 nm        \\
$V$     & DNA velocity          & 100 bp/s      $\approx$ 30 nm/s \\
$R_G$ & Unpackaged genome radius of gyration & 300 nm\\
\hline
\end{tabular}
\caption{\label{tab:parameters}Physical numbers for DNA packing in
  {\phiphage}, taken from~\citet{tao98}.}
\end{center}
\end{table}

First, as DNA is pushed into the capsid, equal and opposite forces pull
both
the unpackaged genome and the capsid through the surrounding fluid.
Although
the (smaller) capsid certainly moves more quickly than the genome, as an
upper
bound, we assume each to move at the full translocation velocity $V$.
Furthermore,
an upper bound can be obtained using the Stokes drag $F = 6\pi \mu R_b V$
on a solid sphere of radius $R_b$ bounding each.  Using $R_b \sim R$ for
the capsid
and $R_b = R_G$ for the genome, we obtain
\be
F_{{\rm cap}} \approx 2 \times 10^{-5}\, \pN\,\,\,{\rm and}\,\,\,F_{{\rm
gen}}\approx 2 \times 10^{-4}\, \pN,
\ee
giving power dissipation
\be
U_{{\rm cap}} \approx 1.5 \times 10^{-4}\, k_B T/\s \,\,\,{\rm
and}\,\,\,U_{{\rm gen}}\approx 1.5 \times 10^{-3}\, k_B T/\s.
\ee

Second, to estimate the dissipation within the sheath, we consider the
fluid
drag on a DNA molecule (modelled as a cylinder) of radius $d$ moving at
velocity $V$ through a cylindrical sheath of inner radius $D$ and length
$X$, into or out of a viral
capsid.  The fluid between the DNA and the sheath is assumed to obey the
Stokes equations
subject to the no-slip boundary conditions.  Packaging or ejecting DNA
requires an equal volume of fluid to be expelled from or injected into the capsid, which can occur
either through pores in the capsid, or back through the sheath.  Below we treat the fluid
dissipation in both cases.

This is a textbook problem in fluid mechanics~\cite{landau87},
giving a fluid velocity profile (with radial coordinate $r$)
\be
u = \frac{V}{\ln d/D} \ln (r/D) + \frac{\Delta P}{4 \mu X}
\left(r^2-d^2 - \frac{(D^2-d^2)}{\ln D/d} \ln r/d\right).
\label{eq:sheathflow}
\ee
The pressure $\Delta P$ depends on the nature of the capsid.
If the capsid is impermeable to water, the volume of DNA $\pi d^2 V$
entering the capsid must exactly equal the fluid volume $2\pi \int_d^D u r
dr$ leaving
the capsid, which gives
\be
\Delta P = \frac{4 \mu V X}{((d^2 + D^2)\ln(D/d)+d^2 - D^2)},
\ee
of order $10^2$ Pa.
The total power dissipated comes from shear stress on the DNA,
$U_{{\rm shear}} = 2 \pi d X V \partial_r u|_{r=d}$ and from $p-V$ work,
driving a volume flux of DNA against a pressure $\Delta P$,
$U_{{\rm pressure}} = \pi d^2 V \Delta P$.  These two give a total
dissipation
\be
U_{{\rm sheath}} = 2 \pi \mu V^2 X\frac{D^2+d^2}{d^2-D^2 + (d^2+D^2)\ln
D/d} \approx 10^{-2} k_B T/\s,
\ee
which represents an upper bound for dissipation within the sheath.

If, as we assume below, it is easier for the fluid to flow through capsid
pores than through the sheath, the `backflow' in (\ref{eq:sheathflow}), 
proportional to $\Delta P$, disappears.  In
that case, only shear stresses occur and give a total dissipation
\be
U_{{\rm sheath}}' = 2\pi X \frac{\mu V^2}{\ln (D/d)} \sim 3\times 10^{-4}
k_B T/\s.
\ee

Finally, we estimate the power dissipated for fluid extruded through
capsid pores, rather than the sheath.  These pores occur at $N$
symmetry points on the capsid shell.  A crude estimate for the dissipation
through each is obtained by assuming Poiseuille flow through each pore.
Entrance and exit effects contribute, at most, a term of comparable
magnitude.  The flow rate through $N$ pores due to a pressure $\Delta P$
inside the capsid is
\be
Q_p \sim \frac{N \Delta P \pi R_p^4}{8\mu t_c}.
\ee
Requiring the DNA volume entering the capsid $\pi d^2 V$ to equal the
fluid flux $Q_p$ out of the capsid determines $\Delta P$ to be
\be
\Delta P \sim \frac{8\mu t_c d^2 V}{N R_p^4},
\ee
giving rise to a viscous dissipation
\be
U_p = \Delta P \pi d^2 V \sim \frac{8\pi \mu t_c d^4 V^2}{N R_p^4}\sim
10^{-7} k_B T/\s.
\ee
Since the dissipation through the pores is so much lower than that through
the sheath, one expects most fluid extrusion to occur primarily through
the pores.

All of these effects are negligible in comparison with the power supplied by 
the motor, which consumes roughly $\sim 10 k_BT$ per 2 base pairs,
giving $W_{\rm motor} \sim 400 k_BT$/s.  

Lastly, we note that if viscous dissipation is the primary damping mechanism during ejection through the sheath,
the DNA ejection velocity can be obtained using the above results.  If fluid must 
flow through the sheath to replace the volume lost by ejecting DNA, the ejection rate 
$V_e \sim 6\mu$m/s per atmosphere of applied pressure; whereas if fluid can flow freely through capsid
pores, higher ejection rates $V_e \sim 300\mu$m/s, per atmosphere
of applied pressure.  Under the first assumption, 10-100kbp genomes (3-30 $\mu$m) would 
be ejected in less than a second, and fifty times faster under the second. 

\subsection{Capsid Mechanics}
In all the calculations above we have assumed the capsid to be rigid. The
capsid is actually a deformable object and it undergoes significant changes in
shape and size during the DNA packaging process. These inelastic changes to the
capsid geometry occuring at the early stages of packing are referred
to in literature under the general heading of \emph{capsid maturation}
or \emph{prohead expansion}~\citep{black88, lata00}. The exact mechanism behind
these changes are not completely understood and are probably not due
to the forces exerted by the DNA~\citep{black88}. In this section we
will be concerned with the later stages of packing after maturation in
an effort to understand the interplay between the forces exerted by
the packaged DNA and the elastic deformation of the mature capsid. In
what follows we treat the capsid as an elastic shell to obtain estimates
about the maximum internal pressure it can sustain. We also examine
the effect of
capsid elasticity on the packaging forces.

The  mature capsid of bacteriophages such as $\lambda, T7 \mbox{ and } HK97$  is made up of copies of a few proteins arranged on an
icosahedral shell. For the purposes of this analysis we model the
capsid as an elastic sphere and assume that the protein sub-units making up the capsids undergo only small
deformations so that the elastic energy stored in a capsid expanded to a
radius $\Rout$ is given by
\begin{equation}
 \Gcap(\Rout) = \kappa(\Rout-R_{0})^{2} \,,
\end{equation}
where $R_{0}$ is the equilibrium radius of an empty capsid and
$\kappa$ is a constant measuring the capsid stiffness. An example
of such an energy emerges from the simulations of \citet{tama02}
who perturb the positions of atoms in the capsid of a plant virus
(CCMV) and measure its energy as a function of radius. They fit a
quadratic polynomial in $\Rout$ to the energy obtained from their
simulations and find
\begin{equation}\label{eq:billi}
 \kappa = 5.0\times10^5 \pN/\nm \,.
\end{equation}
The capsid can be described as a
thin shell of radius $\Rout$.  In the presence of an internal pressure
$p$, the free energy is minimized when
\begin{equation}
-4\pi\Rout^2 p + 2\kappa(\Rout-R_0) = 0\,,
\end{equation}
which implies
\begin{equation}
\Rout-R_0 = {2\pi\Rout^2 p \over \kappa} \,.
\end{equation}


Using eqn.(\ref{eq:billi}) and parameters from a typical
phage capsid, $p=40 \atm$ and $\Rout = 30 \nm$, we find
\begin{equation}
\Rout - R_0 = 4.6\times 10^{-2} \,\nm \,.
\end{equation}
This change is negligible in
comparison to $\Rout$ implying that a phage capsid can be treated
as a rigid shell.

The analysis given above estimates the deformation of the capsid in response
to the forces exerted by the packaged DNA. It is also of interest to
determine the maximum pressure that a capsid can sustain particularly
in view of the osmotic shock experiments on these systems~\citep{cordova03}. We approach this
problem by modeling the capsid as an assemblage of proteins
interacting through weak forces such as van der Waal's forces and
hydrogen bonds. We note that capsids have thin walls compared to their
diameter. For example, the capsid of {\phiphage} is about $1.5$nm
thick while its linear dimensions are of the order of $40$-$50$nm
\cite[see][for data on {\phiphage}]{tao98}. We use these ideas in
conjunction with a coarse-grained model for the cohesive
energies between protein subunits~\citep{Reddy01}
and estimate the maximum
pressure sustainable for a capsid to be in excess of $100 \rm atm$.
The details of the calculations can be found in~\citet{purohit03}.


\section{The DNA Ejection Process}\label{ejection}

We saw in the previous sections that packaged bacteriophage capsids are pressurized
with pressures as high as $60 \atm$. This has led to the speculation that
the high pressure in the bacteriophage provides the driving force for DNA
ejection into the host cell \citep{smith01,kindt01,tzlil03}.  In this section,
we examine the feasibility of this hypothesis. Specifically, we show
that internal forces explain the results in
\citet{evilevitch03}  on inhibition of DNA ejection from phage
$\lambda$ and allows us to make predictions for
bacteriophages with varying genome lengths.

Any experiment in which we can control the amount of DNA ejected from a
bacteriophage by the application of external pressure can help us
understand whether internal forces drive ejection. In a recent
experiment by \citet{evilevitch03}, {\lamphage} bacteriophage were
coerced into ejecting their DNA \emph{in vitro} with the help of a
protein called LamB or maltoporin. This protein, found on the outer
membrane of {\it E. coli} is the natural receptor for {\lamphage}.
When the phage binds to this protein it ejects its cargo of DNA.  The
DNA was ejected into solutions of polyethylene glycol 8000 (PEG) of
various concentrations, which applied known osmotic pressures to the
capsid.  \citeauthor{evilevitch03} found that osmotic pressures of 20~atm
were sufficient to prevent any DNA from leaving the
capsid, whereas DNA was partially ejected at lower external
pressures. 

We can use our model for the forces associated with the packaged
DNA to analyze the experiments of \cite{evilevitch03}. To that end we
consider the system of bacteriophage, ejected DNA and the PEG solution at
equilibrium. The energetics of the DNA inside the phage capsid remains the
same as discussed earlier in this paper. The extra feature we add is the free
energy of the ejected DNA-PEG system. Since the persistence length of
the DNA (50nm) is much larger than the persistence length of the PEG
molecule ($\approx$ 1~nm) we model the insertion of the DNA inside the
PEG solution as being equivalent to the insertion of a rigid cylindrical
rod of radius $R$ inside a solution which exerts osmotic pressure
$\Pi_0$ on the rod. This problem has been studied
by~\citet{castelnovo03}, who estimate the work of insertion of rigid rod (DNA)
into a polymer (PEG 8000) solution as a combination of pressure-volume work,
energy associated with creating new surfaces and the entropic effects
associated with polymer in solution. Here, however,  we resort to a simple
approximation where we retain only the term associated with the
pressure-volume work. Hence the work expended to insert length $L_{0} - L$
of DNA of radius $R$ into the PEG solution is given by
\bea
w(L_{0} - L) = \Pi_{0}(L_{0}-L)\pi R^2.
\eea
The total free energy of the system is the sum of the free energy of
the
DNA inside the phage capsid and the work of insertion and is given by
%
\be
\Gtot(d_{s}(L),L) + \Pi_{0} (L_{0}-L)\pi R^2 .
\label{Gtotal}
\ee
We already know that the free energy of the DNA inside the capsid
depends on the
parameters $F_{0}$ and $c$ which in turn depend on the ionic strength of the
buffer used in the ejection reaction. The experiment by
\citet{evilevitch03} involved a buffer of 10~mM $\rm MgSO_{4}$.
Thus we can use the empirical values $c=0.30$~nm and
$F_0= 1.2\times 10^{4}$~pN/nm$^2$
for {\mgion} solutions as determined earlier.  We will use
these values for all the fits to the experimental data and for further
predictions. \\

\begin{figure}
\centering
\includegraphics{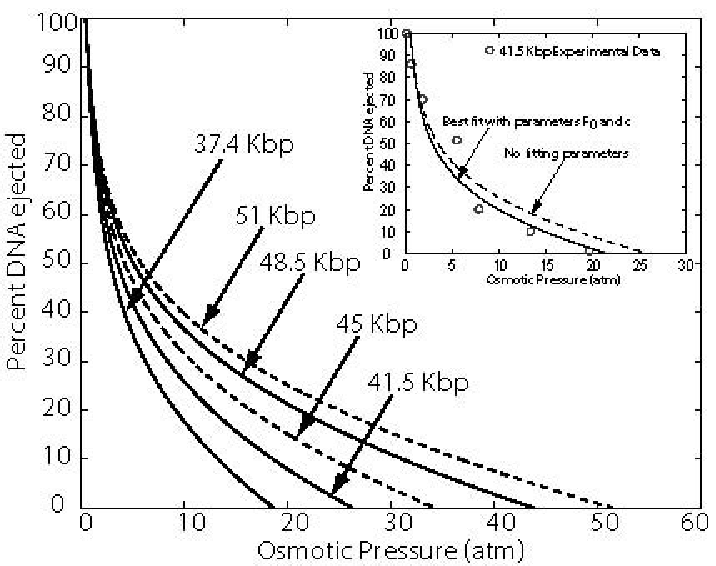}
\caption{Fractional DNA ejection in $\lambda$-phage as a function of osmotic
  pressure corresponding to experimental conditions
  in~\cite{evilevitch03} ($10\rm mM$ $\rm MgSO_{4}$). The inset shows
  the best visual fit to the experimental data of~\cite{evilevitch03}
  for {\lamphage} with genome size $41.5$ kbp using parameters $F_{0}$
  and $c$, and also a parameter free curve obtained from the data
  of~\citet{rau84}. We see a good match between the experimental
  findings and the theoretical results. Also shown are the predictions
  for ejection behaviour of $\lambda$-phage for genome sizes of 37.4
  Kbp, 45 Kbp, 48.5 Kbp and 51 Kbp under similar experimental conditions.}
\label{plot_length}
\end{figure}

\begin{figure}
\centering
\includegraphics{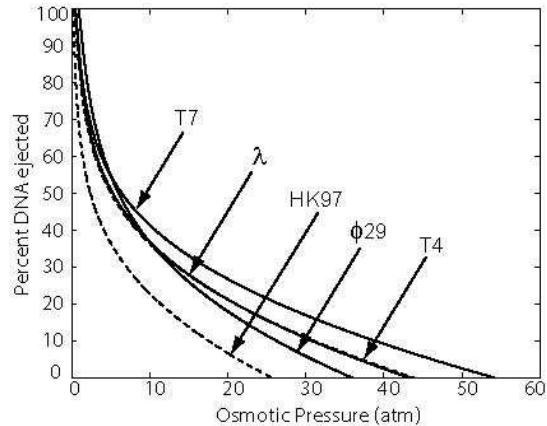}
\caption{DNA ejection as a function of osmotic pressure for various
  wild-type species of bacteriophage for ionic conditions similar to
  the experiment of~\cite{evilevitch03} ($10 \rm mM$ $\rm MgSO_{4}$). The lines show the DNA ejection
  behavior for T4, HK97, {\phiphage}, T7 and  $\lambda$.
  Osmotic pressures as high as 35-55atm are required to
  inhibit ejection in these bacteriophage.}\label{plot_virus}
\end{figure}

In order to find find $L_{0}-L$, the ejected length, we
need to minimize the free energy $\Gtot$ with respect to $L$. Differentiating
eqn.(\ref{Gtotal}) we get, at equilibrium,
\bea \label{eq}
{\partial \Gtot(d_{s}(L),L)\over \partial L}- \Pi_{o}\pi R^{2} = 0 .
\eea
The first term in the above expression is merely the resisting force
$F(d(L),L)$ derived earlier. Consequently equation(\ref {eq}) becomes
\bea
F(d_{s}(L),L) = \Pi_{o} \pi R^{2} .
\eea
This equation is solved for $L$ for several values of $\Pi_{0}$ and the
results for the percentage of DNA ejected as a function of
osmotic pressure are given in Figure~\ref{plot_length}, for different total
DNA lengths $L_0$. Experiments
are currently underway to examine the extent to which the predictions
described here are borne out experimentally.
Figure~\ref{plot_virus} shows how the model applies to the ejection
behavior of other phage under the same solvent conditions as in~\citet{evilevitch03}.


\section{Conclusions}\label{conclusion}

This paper addresses  physical processes in the viral life cycle
through a quantitative framework based on insights from structural
biology, single molecule biophysics, electron microscopy and solution
biochemistry. The models were motivated by specific experiments on
{\phiphage}~\citep{smith01} and {\lamphage}~\citep{evilevitch03} but
their applicability extends to all dsDNA bacteriophages. In fact, we
use our models to predict important features of the packaging and
ejection processes in phages other than {\phiphage} and~{\lamphage}.



The key predictions arising from the modeling efforts described here are:
\begin{itemize}

\item{{\it Dependence of Forces and Spacing on Ionic Strength.}  As shown in figs. 4,5,7 and 10, there is a strong dependence of both the
spacing of the packaged DNA as well as the forces that build up due to packing on the ionic conditions during the packaging reaction.   We
suggest systematic experiments to explore these effects.}

\item{{\it Dependence of Forces and Spacing on Phage Identity.} As shown in figs. 7 and 8, we find a systematic and strong dependence of the
packing forces on the particular phage species of interest.   In particular, had the experiment of Smith {\it et al.} been carried out in
phage T7, we predict a maximum packing force in excess of 100pN, rather than the 57pN found in $\phi$29.  
A simple parameter for developing intuition concerning the forces associated with different phage is $\rho_{pack}$,
the ratio of the volume of the genome to the volume of the capsid.  }

\item{{\it Force Steps During Packaging.}  One of the weakest points of the analysis described in this paper is the uncertainties that
attend the particular structural arrangements of the DNA on the way to the fully packaged state.  In particular, we have {\it assumed} a
sequence of structural states which are all of the inverse spool form and one consequence of this structural picture which might be testable
is the presence of steps in both the DNA spacing and forces as shown in fig. 9.}

\item{{\it Dependence of ejection inhibition on genome length, virus
    type and solution conditions. } The beautiful experiments of
    \citet{evilevitch03} provide a direct window on the forces
    associated with the packaged
    DNA. Fig.~\ref{plot_length}~and~\ref{plot_virus} represent a wide
    range of parameter-free prediction for the fractional ejection
    inhibition that should be seen in such experiments.}

\end{itemize}

Finally, a cautionary note.  Experiments have shown that some
bacteriophage, such as T7~\citep{molineux01} may rely on a
different mechanism for delivering their genome into the host
cell. This is a possibility worthy of further exploration, but we
emphasize that viruses may use many different methods or
combinations thereof to propagate themselves.

{\it Acknowledgments.}  We are grateful to  Alex Evilevitch, Chuck Knobler, Jon Widom, Bill Gelbart, Andy
Spakowitz, Zhen-Gang Wang, Ken Dill, Carlos Bustamante, Larry Friedman,
Jack Johnson, Paul Wiggins, Steve Williams, Wayne Falk, Adrian
Parsegian, Alasdair Steven, Florence Tama, Vijay Reddy, Charlie
Brooks, Peter Privelege, Steve Harvey, Ian Molineux and Steve Quake.
Rob Phillips and Prashant Purohit  acknowledge support of the NSF
through grant number CMS-0301657, the NSF supported CIMMS center and the support of the Keck Foundation.
Jan\'{e} Kondev is supported by the NSF under grant number DMR-9984471, and is a
Cottrell Scholar of Research Corporation.   PG was supported by an NSF
graduate research fellowship.

\bibliographystyle{elsart-harv}
\bibliography{jmbpaper}

\begin{thebibliography}{70}
\expandafter\ifx\csname natexlab\endcsname\relax\def\natexlab#1{#1}\fi
\expandafter\ifx\csname url\endcsname\relax
  \def\url#1{\texttt{#1}}\fi
\expandafter\ifx\csname urlprefix\endcsname\relax\def\urlprefix{URL }\fi

\bibitem[{Alberts et~al.(1997)Alberts, Bray, Johnson, Lewis, Raff, Roberts, and
  Walter}]{alberts97}
Alberts, B., Bray, D., Johnson, A., Lewis, J., Raff, M., Roberts, K., Walter,
  P., Feb 1997. Essential Cell Biology. Garland Publishing, New York, the issue
  of the relative dimensions of DNA and the regions in which it is packed is
  also explored in Austin {\it et al.} (Feb. 1997), {\it Phys. Today}.

\bibitem[{Arsuaga et~al.(2002)Arsuaga, Tan-Z., Vazquez, Sumners, and
  Harvey}]{arsuaga02}
Arsuaga, J., Tan-Z., R.~K., Vazquez, M., Sumners, D.~W., Harvey, S.~C., 2002.
  Investigation of viral {DNA} packaging using molecular mechanics models.
  Biophys.\ Chem. 101--102, 475--484.

\bibitem[{Baker et~al.(1999)Baker, Olson, and Fuller}]{baker99}
Baker, T.~S., Olson, N.~H., Fuller, S.~D., 1999. Adding the third dimension to
  virus life cycles: Three-dimensional reconstruction of icosahedral viruses
  from cryo-electron micrographs. Microbiol.\ and\ Mol.\ Bio.\ Rev. 63, 862.

\bibitem[{Bednar et~al.(1995)Bednar, Furrer, Katritch, Stasiak, Dubochet, and
  Stasiak}]{bednar95}
Bednar, J., Furrer, P., Katritch, V., Stasiak, A.~Z., Dubochet, J., Stasiak,
  A., 1995. Determinition of {DNA} persistence length by cryo-electron
  microscopy. seperation of static and dynamic contributions to the apparent
  persistence length of {DNA}. J.~Mol.\ Biol. 254, 579--594.

\bibitem[{Black(1988)}]{black88}
Black, L., 1988. {DNA} packaging in {dsDNA} bacteriophage. In: R.Calendar
  (Ed.), The Bacteriophages. Vol.~2. Plenum Press, Ch.~5, pp. 321--363.

\bibitem[{Bohm et~al.(2001)Bohm, Lambert, Frangakis, Letellier, Baumeister, and
  Rigaud}]{bohm01}
Bohm, J., Lambert, O., Frangakis, A., Letellier, L., Baumeister, W., Rigaud,
  J., 2001. {FhuA}-mediated phage genome transfer into liposomes: A
  cryo-electron tomography study. Curr.\ Biol., 1168--1175.

\bibitem[{Booy et~al.(1991)Booy, Newcomb, Trus, Brown, Baker, and
  Steven}]{booy91}
Booy, F.~P., Newcomb, W.~W., Trus, B.~L., Brown, J.~C., Baker, T.~S., Steven,
  A.~C., 1991. Liquid-crystalline, phage-like packing of encapsidated {DNA} in
  herpes-simplex virus. Cell 64~(5), 1007--1015.

\bibitem[{Castelnovo et~al.(2003)Castelnovo, Bowles, Reiss, and
  Gelbart}]{castelnovo03}
Castelnovo, M., Bowles, R.~K., Reiss, H., Gelbart, W.~M., 2003. Osmotic force
  resisting chain insertion in a colloidal suspension. Euro.\ Phys.\ J.\ E 10,
  191.

\bibitem[{Cerritelli et~al.(1997)Cerritelli, Cheng, A.~H.~Rosenberg, Booy, and
  Steven}]{cerritelli97}
Cerritelli, M.~E., Cheng, N., A.~H.~Rosenberg, M. E.~P., Booy, F.~P., Steven,
  A.~C., 1997. Encapsidated conformation of bacteriophage {T7} {DNA}. Cell 91,
  271.

\bibitem[{Cloutier and Widom(2004)}]{cloutier04}
Cloutier, T., Widom, J., 2004. Spontaneous sharp bending of double-stranded
  {DNA}. Mol.\ Cell. 14, 355--362.

\bibitem[{Cordova et~al.(2003)Cordova, Deserno, Gelbart, and
  Ben-Shaul}]{cordova03}
Cordova, A., Deserno, M., Gelbart, W.~M., Ben-Shaul, A., 2003. Osmotic shock
  and the strength of viral capsids. Biophys. J. 85, 70--74.

\bibitem[{de~Gennes(1979)}]{gennes79}
de~Gennes, P.~G., 1979. Scaling concepts in polymer physics. Cornell University
  Press, Ithaca, NY.

\bibitem[{de~Vries(2001)}]{vries01}
de~Vries, R., 2001. Flexible polymer-induced condensation and bundle formation
  of {DNA} and {F}-actin filaments. Biophys.~J. 80, 1186.

\bibitem[{Doi(1996)}]{doibook}
Doi, M., Feb. 1996. Introduction to polymer physics. Clarendon Press.

\bibitem[{Dokland and Murialdo(1993)}]{dokland93}
Dokland, T., Murialdo, H., 1993. Structural transitions during maturation of
  bacteriophage lambda capsids. J.~Mol.\ Biol. 223.

\bibitem[{Earnshaw and Casjens(1980)}]{earnshaw80}
Earnshaw, W.~C., Casjens, S.~R., 1980. {DNA} packaging by double-stranded {DNA}
  bacteriophages. Cell 21, 319--331.

\bibitem[{Earnshaw and Harrison(1977)}]{earnshaw77}
Earnshaw, W.~C., Harrison, S.~C., 1977. {DNA} arrangement in isometric phage
  heads. Nature 268, 598--602.

\bibitem[{Echols(2001)}]{echols01}
Echols, H., 2001. Operators and promoters: {The} story of molecular biology and
  its creators. University of California Press, Berkeley,California.

\bibitem[{Endy et~al.(1997)Endy, Kong, and Yin}]{endy97}
Endy, D., Kong, D., Yin, J., 1997. Intracellular kinetics of a growing virus:
  {A} genetically structured simulation for simulation for the growth of
  bacteriophage {T7}. Biotech. Bioeng. 55, 375--389.

\bibitem[{Evans(2001)}]{evans01}
Evans, E., 2001. Probing the relation between force--lifetime--and chemistry in
  single molecular bonds. Ann.\ Rev.\ Biophys.\ Biomol.\ Struct. 30, 105.

\bibitem[{Evilevitch et~al.(2004)Evilevitch, Castelnovo, Knobler, and
  Gelbart}]{evilevitch04}
Evilevitch, A., Castelnovo, M., Knobler, C., Gelbart, W., 2004. Measuring the
  force ejecting {DNA} from phage. J. \ Phys. \ Chem. \ B 108~(21), 6838--6843.

\bibitem[{Evilevitch et~al.(2003)Evilevitch, Lavelle, Knobler, Raspaud, and
  Gelbart}]{evilevitch03}
Evilevitch, A., Lavelle, L., Knobler, C.~M., Raspaud, E., Gelbart, W.~M., 2003.
  Osmotic pressure inhibition of {DNA} ejection from phage. Proc.\ Natl.\
  Acad.\ Sci.\ USA 100, 9292.

\bibitem[{Feiss et~al.(1977)Feiss, Fisher, Crayton, and Egner}]{feiss77}
Feiss, M., Fisher, R.~A., Crayton, M.~A., Egner, C., 1977. Packaging of the
  bacteriophage $\lambda$ chromosome: Effect of chromosome length. Virology 77,
  281--293.

\bibitem[{Flint et~al.(2000)Flint, Enquist, Krug, Racaniello, and
  Skalka}]{flint}
Flint, S.~J., Enquist, L.~W., Krug, R.~M., Racaniello, V.~R., Skalka, A.~M.,
  2000. Principles of Virology. ASM Press, Washington, DC.

\bibitem[{Gelbart et~al.(2000)Gelbart, Bruinsma, Pincus, and
  Parsegian}]{gelbart00}
Gelbart, W.~M., Bruinsma, R.~F., Pincus, P.~A., Parsegian, V.~A., Sep. 2000.
  {DNA} inspired electrostatics. Physics\ Today, 38--44.

\bibitem[{Hershey and Chase(1952)}]{hershey52}
Hershey, A., Chase, M., 1952. Independent functions of viral protein and
  nucleic acid in growth of bacteriophage. J. Gen. Physiol. 36~(1), 39--56.

\bibitem[{Iwasaki et~al.(2000)Iwasaki, Trus, Wingfield, Cheng, Campusano, Rao,
  and Steven}]{iwasaki00}
Iwasaki, K., Trus, B.~L., Wingfield, P.~T., Cheng, N., Campusano, G., Rao,
  V.~B., Steven, A.~C., 2000. Molecular architecture of bacteriophage {T4}
  capsid: Vertex structure and bimodal binding of the stabilizing accessory
  protein, {Soc}. Virology 271, 321--333.

\bibitem[{Kanamaru et~al.(2002)Kanamaru, Leiman, Kostyuchenko, Chipman,
  Mesyanzhinov, Arisaka, and Rossmann}]{kanamaru02}
Kanamaru, S., Leiman, P.~G., Kostyuchenko, V.~A., Chipman, P.~R., Mesyanzhinov,
  V.~M., Arisaka, F., Rossmann, M.~G., 2002. Structure of the cell-puncturing
  device of bacteriophage {T4}. Nature 415, 553.

\bibitem[{Kindt et~al.(2001)Kindt, Tzlil, Ben-Shaul, and Gelbart}]{kindt01}
Kindt, J., Tzlil, S., Ben-Shaul, A., Gelbart, W., 2001. {DNA} packaging and
  ejection forces in bacteriophage. Proc.\ Natl.\ Acad.\ Sci.\ USA 98, 13671.

\bibitem[{Klug and Ortiz(2003)}]{klug03}
Klug, W.~S., Ortiz, M., 2003. A director field model of {DNA} packaging in
  viral capsids. J.~Mech.\ Phys.\ Sol. 51~(10), 1815--1847.

\bibitem[{{La Scola} et~al.(2003){La Scola}, Audic, Robert, Jungang,
  de~Lamballerie, Drancourt, Birtles, and J.-M.~Claverie}]{lascola03}
{La Scola}, B., Audic, S., Robert, C., Jungang, L., de~Lamballerie, X.,
  Drancourt, M., Birtles, R., J.-M.~Claverie, D.~R., 2003. A giant virus in
  amoebae. Science 299, 2033.

\bibitem[{LaMarque et~al.(2004)LaMarque, Le, and Harvey}]{lamarque04}
LaMarque, J., Le, T., Harvey, S., 2004. Packaging double-helical {DNA} into
  viral capsids. Biopolymers 73, 348--355.

\bibitem[{Landau and Lifshitz(1987)}]{landau87}
Landau, L.~D., Lifshitz, E.~M., 1987. Fluid Mechanics, 2nd Edition.
  Butterworth-Heinemann.

\bibitem[{Lata et~al.(2000)Lata, Conway, Cheng, Duda, Hendrix, Wikoff, Johnson,
  Tsuruta, and Steven}]{lata00}
Lata, R., Conway, J.~F., Cheng, N., Duda, R.~L., Hendrix, R.~W., Wikoff, W.~R.,
  Johnson, J.~E., Tsuruta, H., Steven, A.~C., 2000. Maturation dynamics of a
  viral capsid: Visualization of transitional intermediate states. Cell 100,
  253--263.

\bibitem[{Leiman et~al.(2003)Leiman, Kanamaru, Mesayanzhinov, Arisaka, and
  Rossmann}]{leiman03}
Leiman, P.~G., Kanamaru, S., Mesayanzhinov, V.~V., Arisaka, F., Rossmann,
  M.~G., 2003. Structure and morphogenesis of bacteriophage {T4}. Cell. Mol.
  Life Sci. 60, 2356--2370.

\bibitem[{Letellier et~al.(2004)Letellier, Boulanger, Plancon, Jacquot, and
  Santamaria}]{letellier04}
Letellier, L., Boulanger, P., Plancon, L., Jacquot, P., Santamaria, M., 2004.
  Main features of tailed phage, host recognition and {DNA} uptake. Front.
  Biosci. 9, 1228--1239.

\bibitem[{Lyubartsev and Nordenskiold(1995)}]{lyubartsev95}
Lyubartsev, A.~P., Nordenskiold, L., 1995. A {Monte Carlo} simulation study of
  ion distribution and osmotic pressure in hexagonally oriented {DNA}. Phys.
  Rev. E 99, 10373.

\bibitem[{Marenduzzo and Micheletti(2003)}]{marenduzzo03}
Marenduzzo, D., Micheletti, C., 2003. Thermodynamics of {DNA} packing inside a
  viral capsid: {The} role of {DNA} intrinsic thickness. J.~Mol.\ Biol.
  330~(3), 485--492.

\bibitem[{Martin et~al.(2001)Martin, Burnett, Hass, Heinkel, Rutten, Fuller,
  Buthcher, and Bamford}]{martin01}
Martin, S.~C., Burnett, R.~M., Hass, F., Heinkel, R., Rutten, T., Fuller,
  S.~D., Buthcher, S.~J., Bamford, D.~H., 2001. Combined {EM/X}-ray imaging
  yields a quasi-atomic model of the adenovirus related bacteriophage {PRD1}
  and shows key capsid and membrane interactions. Structure 9, 917--930.

\bibitem[{Molineux(2001)}]{molineux01}
Molineux, I.~J., 2001. No syringes please, ejection of phage {T7} {DNA} from
  the virion is enzyme driven. Mol.\ Microbiol. 40, 1--8.

\bibitem[{{National Center For Biotechnology
  Information}(2004)}]{entrezgenomes}
{National Center For Biotechnology Information}, 2004. Entrez genomes. Online
  at {\tt http://www.ncbi.nlm.nih.gov/}.

\bibitem[{Neidhardt(1996)}]{neidhardt96}
Neidhardt, F. (Ed.), 1996. Escherichia Coli and Salmonella Typhimurium. ASM
  Press.

\bibitem[{Nelson(2003)}]{nelson03}
Nelson, P., 2003. Biological Physics: Energy, Information, Life. W. H. Freeman
  \& Co.

\bibitem[{Novick and Baldeschwieler(1988)}]{novick88}
Novick, S., Baldeschwieler, J., 1988. Fluorescence measurement of the kinetics
  of {DNA} injection by bacteriophage $\lambda$ into liposomes. Biochemistry
  27, 7919--7924.

\bibitem[{Odijk(1998)}]{odijk98}
Odijk, T., 1998. Hexagonally packed {DNA} within bacteriophage {T7} stabilized
  by curvature stress. Biophys.~J. 75, 1223.

\bibitem[{Odijk(2003)}]{odijk03}
Odijk, T., 2003. Statics and dynamics of condensed {DNA} within phages and
  globules, submitted.

\bibitem[{Odijk and Slok(2003)}]{odijk03b}
Odijk, T., Slok, F., 2003. Nonuniform {Donnan} equilibrium within
  bacteriophages packed with {DNA}. J.~Phys.\ Chem.~B 107~(32), 8074--8077.

\bibitem[{Olson et~al.(2001)Olson, Gingery, Eiserling, and Baker}]{olson01}
Olson, N.~H., Gingery, M., Eiserling, F.~A., Baker, T.~S., 2001. The structure
  of isometric capsids of bacteriophage {T4}. Virology 279, 385.

\bibitem[{Parsegian et~al.(1986)Parsegian, Rand, Fuller, and Rau}]{parsegian86}
Parsegian, V.~A., Rand, R.~P., Fuller, N.~L., Rau, D.~C., 1986. Osmotic stress
  for the direct measurement of intermolecular forces. Meth.\ Enzym. 127, 400.

\bibitem[{Phillips(2001)}]{phillips01}
Phillips, R., 2001. Crystals, Defects and Microstructures. Cambridge University
  Press.

\bibitem[{Ptashne(2004)}]{ptashne04}
Ptashne, M., 2004. Genetic switch: Phage lambda revisited. Cold Spring Harbor
  Laboratory.

\bibitem[{Purohit et~al.(2003{\natexlab{a}})Purohit, Kondev, and
  Phillips}]{purohit03b}
Purohit, P.~K., Kondev, J., Phillips, R., 2003{\natexlab{a}}. Force steps in
  viral {DNA} packaging ? J.~Mech.\ Phys.\ Sol. 51~(11), 2239--2257.

\bibitem[{Purohit et~al.(2003{\natexlab{b}})Purohit, Kondev, and
  Phillips}]{purohit03}
Purohit, P.~K., Kondev, J., Phillips, R., 2003{\natexlab{b}}. Mechanics of
  {DNA} packaging in viruses. Proc.\ Natl.\ Acad.\ Sci.\ USA 100~(6),
  3173--3178.

\bibitem[{Raspaud et~al.(1998)Raspaud, de~la Cruz, Sikorav, and
  Livolant}]{raspaud98}
Raspaud, E., de~la Cruz, M.~O., Sikorav, J.~L., Livolant, F., 1998.
  Precipitation of {DNA} by polyamines: A polyelectrolyte behavior. Biophys.~J.
  74~(1), 381--393.

\bibitem[{Rau et~al.(1984)Rau, Lee, and Parsegian}]{rau84}
Rau, D.~C., Lee, B., Parsegian, V.~A., 1984. Measurement of the repulsive force
  between polyelectrolyte molecules in ionic solution: Hydration forces between
  parallel {DNA} double helices. Proc.\ Natl.\ Acad.\ Sci.\ USA 81, 2621.

\bibitem[{Rau and Parsegian(1992)}]{rau92}
Rau, D.~C., Parsegian, V.~A., 1992. Direct measurement of the intermolecular
  forces between counterion-condensed {DNA} double helices. Biophys.~J. 61,
  246.

\bibitem[{Reddy et~al.(2001)Reddy, Natarajan, Okerberg, Li, Damodaran, Morton,
  and Johnson}]{Reddy01}
Reddy, V.~S., Natarajan, P., Okerberg, B., Li, K., Damodaran, K.~V., Morton,
  R.~T., Johnson, C. L. B. J.~E., 2001. Virus particle explorer ({VIPER}), a
  website for virus capsid structures and their computational analyses.
  Virology 75, 11943, the Viper website can be found at {\tt
  http://mmtsb.scripps.edu/viper/viper.html}.

\bibitem[{Richards et~al.(1973)Richards, Williams, and Calendar}]{richards73}
Richards, K.~E., Williams, R.~C., Calendar, R., 1973. Mode of {DNA} packing
  within bacteriophage heads. J.~Mol.\ Biol. 78, 255.

\bibitem[{Riemer and Bloomfield(1978)}]{riemer78}
Riemer, S.~C., Bloomfield, V.~A., 1978. Packaging of {DNA} in bacteriophage
  heads: Some considerations on energetics. Biopolymers 17, 785.

\bibitem[{Schnitzer et~al.(2000)Schnitzer, Visscher, and Block}]{schnitzer00}
Schnitzer, M.~J., Visscher, K., Block, S.~M., 2000. Force production by single
  kinesin motors. Nature Cell.\ Biol. 2, 718--723.

\bibitem[{Shibata et~al.(1987)Shibata, Fujisawa, and Minagawa}]{shibata87}
Shibata, H., Fujisawa, H., Minagawa, T., 1987. Characterization of the
  bacteriophage {T3} {DNA} packaging reaction in vitro in a defined system.
  J.~Mol.\ Biol. 196, 845--851.

\bibitem[{Simpson et~al.(2000)Simpson, Tao, Leiman, Badasso, He, Jardine,
  Olson, Morais, Grimes, Anderson, Baker, and Rossmann}]{simpson00}
Simpson, A.~A., Tao, Y., Leiman, P.~G., Badasso, M.~O., He, Y., Jardine, P.~J.,
  Olson, N.~H., Morais, M.~C., Grimes, S., Anderson, D.~L., Baker, T.~S.,
  Rossmann, M.~G., 2000. Structure of the bacteriophage $\phi$29 {DNA}
  packaging motor. Nature 408, 745.

\bibitem[{Smith et~al.(2001)Smith, Tans, Smith, Grimes, Anderson, and
  Bustamante}]{smith01}
Smith, D.~E., Tans, S.~J., Smith, S.~B., Grimes, S., Anderson, D.~L.,
  Bustamante, C., 2001. The bacteriophage $\phi$29 portal motor can package
  {DNA} against a large internal force. Nature 413, 748.

\bibitem[{Smith et~al.(1992)Smith, Finzi, and Bustamante}]{smith92}
Smith, S.~B., Finzi, L., Bustamante, C., 1992. Direct mechanical measurements
  of the elasticity of single dna molecules by using magnetic beads. Science
  258, 1122--1126.

\bibitem[{Tama and Brooks(2002)}]{tama02}
Tama, F., Brooks, C.~L., 2002. The mechanism and pathway of {pH} induced
  swelling in cowpea chlorotic mottle virus. J.~Mol.\ Biol. 318, 733--747.

\bibitem[{Tao et~al.(1998)Tao, Olson, Xu, Anderson, Rossmann, and
  Baker}]{tao98}
Tao, Y., Olson, N.~H., Xu, W., Anderson, D.~L., Rossmann, M.~G., Baker, T.~S.,
  1998. Assembly of a tailed bacterial virus and its genome release studied in
  three dimensions. Cell 95, 431.

\bibitem[{Tzlil et~al.(2003)Tzlil, Kindt, Gelbart, and Ben-Shaul}]{tzlil03}
Tzlil, S., Kindt, J., Gelbart, W., Ben-Shaul, A., 2003. Forces and pressures in
  dna packaging and release from viral capsids. Biophys. \ J. 84, 1616--1627.

\bibitem[{Wikoff and Johnson(1999)}]{wikoff99}
Wikoff, W.~R., Johnson, J.~E., 1999. Virus assembly: Imaging a molecular
  machine. Curr.\ Biol. 9, R296.

\bibitem[{{World Health Organization}(2004)}]{whosmallpox}
{World Health Organization}, 2004. Smallpox. Fact sheet online at {\tt
  http://www.who.int/mediacentre/factsheets/smallpox/en/}.

\bibitem[{Young(1992)}]{young92}
Young, R.~Y., 1992. Bacteriophage lysis:mechanism and regulation. Microbiol.
  Rev. 56~(3), 430--481.

\end{thebibliography}

\end{document}